\documentclass[a4paper,fleqn]{cas-dc}

\usepackage[numbers]{natbib}
\usepackage{hyperref}
\usepackage{bookmark}

\def\tsc#1{\csdef{#1}{\textsc{\lowercase{#1}}\xspace}}
\tsc{WGM}
\tsc{QE}
\tsc{EP}
\tsc{PMS}
\tsc{BEC}
\tsc{DE}

\begin{document}
\let\WriteBookmarks\relax
\def\floatpagepagefraction{1}
\def\textpagefraction{.001}

\shorttitle{Electronic structure and effective mass analysis of doped TiO$_2$ (anatase) systems using DFT+$U$}

\shortauthors{A. Raghav \textit{et al.}}

\title [mode = title]{Electronic structure and effective mass analysis of doped TiO$_2$ (anatase) systems using DFT+$U$}                      



%
%

\author[1]{Abhishek Raghav}[
    orcid=0000-0002-5881-1290
]

\cormark[1]


\ead{mwkabr1915@icloud.com}


\credit{Conceptualization of this study, Methodology, Software, Data curation, Writing - Original draft preparation}

\affiliation[1]{organization={School of Information Science, Japan Advanced Institute of Science and Technology (JAIST)},
    city={Nomi},
    postcode={923-1292}, 
    state={Ishikawa},
    country={Japan}}

\affiliation[2]{organization={Research Center for Advanced Computing Infrastructure, Japan Advanced Institute of Science and Technology (JAIST)},
    city={Nomi},
    postcode={923-1292}, 
    state={Ishikawa},
    country={Japan}}

\author[2]{Kenta Hongo}

\credit{Conceptualization, Validation, Writing-review \& editing}

\author[1]{Ryo Maezono}


\credit{Conceptualization, Validation, Writing-review \& editing, Supervision, Project administration}

\affiliation[3]{organization={Materials Engineering, Indian Institute of Technology Gandhinagar (IITGN)},
    city={Gandhinagar},
    postcode={382355}, 
    state={Gujarat},
    country={India}}

\author[3]
{Emila Panda}
\cormark[2]
\credit{Conceptualization, Validation, Writing-review \& editing, Supervision, Project administration}

\ead{emila@iitgn.ac.in}


\cortext[cor1]{Corresponding author}
\cortext[cor2]{Principal corresponding author}



\begin{abstract}
    In this work, electronic structure of several doped TiO$_2$ anatase systems is computed using DFT+$U$.
    Effective masses of charge carriers are also computed to quantify how the dopant atoms perturb the bands of the host anatase material.
    $U$ is computed systematically for all the dopants using the linear response method rather than using fitting procedures to physically known quantities.
    A combination of $d$ and $f$ block elements (Nb, Ta, V, Mo, W, Cr, La, Cu, Co and Ce) are considered as dopants.
    Depending upon the energies of their outer $d$ or $f$ electrons, the dopants are found to form defect states at various positions in the band structure of host anatase system.
    Some dopants like Cr, Mo etc. form mid-gap states, which could reduce transparency.
    Other dopants like Nb, Ta and W are found to have the Fermi levels positioned near the conduction band edge, indicating these systems to exhibit \textit{n}-type conductivity.
    From the effective mass analysis, dopants are found to increase the effective mass of charge carriers and the non-parabolic nature of bands.
    Based on electronic structure and effective mass analysis, Nb, Ta and W are identified to exhibit higher transparency and conductivity as compared to the other dopants considered here.
    The theoretical results presented here, increase our understanding and show the potential of dopants to alter the properties in anatase TiO$_2$.
\end{abstract}


\begin{highlights}
\item Electronic structure of doped TiO$_2$ (anatase) systems is computed using DFT+$U$.
\item $U$ is computed systematically from first principles using the linear response approach.
\item Effective masses of charge carriers are computed for the doped systems.
\item Doped systems for possible TCO application are identified.
\end{highlights}

\begin{keywords}
doped anatase \sep electronic structure \sep effective mass \sep density functional theory \sep linear response method \sep transparent conducting oxide
\end{keywords}

\maketitle

\section{Introduction}
TiO$_2$ (anatase) is an attractive transition metal oxide known for its applications in transparent conducting oxides (TCO), photocatalysts and photovoltaics~\cite{2007HIT,2010HIT,1972FUJ,2012NAK}.  
It has an experimentally measured band gap of 3.2 eV~\cite{1994TAN}, which makes it inherently transparent, but also a poor electrical conductor.
Native defects and dopants can form defect states and can be used to engineer or tune the band structure for specific applications.
This is called as band-gap or band-structure tuning.
For example, anatase thin films when doped with Niobium are found to demonstrate good conductivity, thus making this material suitable for TCO application~\cite{2005HIT}.
There have been several computational and experimental studies to find new and better dopants and to also understand how native defects and dopants~\cite{2020RAG, 2006PHA, 2010MOR, 2021MAN2} affect the electronic structure of the host anatase material. 

Several earlier studies used LDA or GGA to investigate doped anatase systems~\cite{2002UME,2006PHA,2014WAN}. 
These had obvious disadvantages of band gap underestimation and delocalization of excess electrons over the crystal due to the self interaction error. 
Hence, the more recent studies were seen to have used the Hubbard correction (GGA+$U$)~\cite{2013HAN,2012CHE}, hybrid DFT etc. to improve the accuracy of the computation. 
However, to the best of our knowledge, in almost all the studies for doped anatase using DFT+$U$, $U$ has been determined by using fitting to physically known quantities (like band gap) or taken from earlier literature~\cite{2009MOR,2011ARR,2012CHE}, hence limiting the predictive ability of the theory.
Here, we used an alternative way: computing the value of $U$ for all the dopants consistently using the linear response approach developed by Cococcioni \textit{et al.}~\cite{2005COC}.
Note that, this first principles approach does not rely on fitting $U$ value to physically known quantities and thus could lead to better predictions.

Depending on the energy of $d$ electrons, dopants might act as shallow donors, form a mid-gap or acceptor like state~\cite{2008OSO}. 
In the first case, the defect state is formed very close to the conduction band edge, such that the band gap is only marginally affected and electron transition from the defect state to the conduction band can impart \textit{n}-type conductivity to the host material at room temperature. 
Dopant states formed near the conduction band edge might also affect the curvature of the bands, altering the effective mass of the charge carriers. 
In the second case, a deep mid-gap state is formed. 
Formation of such a mid-gap state might adversely affect the TCO application because of the possible electron transition to and from the mid-gap state (hence reducing the transparency). 
In the third case, dopant forms acceptor like state close to the valence band. 
This could lead to \textit{p}-type conductivity in the host material. 
Again here, the dopant state might modify the band curvature, thereby modifying the effective mass of the charge carriers. 
There are some other possibilities as well. 
For example, the dopant atom might form delocalized states in the conduction band, donating excess electrons in the conduction band and imparting a metallic nature to the system. 
Dopant atom might also form a combination of localized mid-gap and delocalized states.

Effective mass of the charge carriers plays a crucial role in inducing the optoelectronic properties in the materials. 
Effective mass analysis has been done for Nb-doped anatase~\cite{2008HIT}, but to the best of our knowledge, for other dopants the reports are quite scarce and this area remains largely unexplored. 
Hence, in this study, we carried out a comprehensive effective mass analysis for the charge carriers for several dopants, to understand how dopants perturb the band curvatures of the host anatase system. 
We believe that this information would be crucial to both experimentalists and computational material scientists alike.

\section{Methodology}
\label{methodology}
\subsection{Computational details}
\label{computational_details}
Spin-polarized DFT+$U$ calculations were performed using the Vienna \textit{Ab initio} Simulation Package (VASP), utilizing the plane-wave basis sets~\cite{1994KRE,1996KRE,1996KRE2}. 
Projector augmented wave type pseudopotentials were used to model the core electrons~\cite{1994BLO}.
The valence configurations considered are listed in Table.~\ref{tab:tabl1}

 \begin{table}[htbp,width=.7\linewidth]
 \caption{Valence configurations of various elements used in this work}
 \label{tab:tabl1}
 \begin{tabular*}{\tblwidth}{@{}lc@{}}
 \toprule
 Element & Valence configuration   \\ \midrule
 Ti      & 3$p^6$ 3$d^3$ 4$s^1$    \\  
 O       & 2$s^2$ 2$p^4$           \\  
 Nb      & 4$p^6$ 5$s^1$ 4$d4$     \\  
 Ta      & 5$p^6$ 6$s^1$ 5$d4$     \\  
 V       & 3$p^6$ 3$d^4$ 4$s^1$    \\  
 Mo      & 4$p^6$ 5$s^1$ 4$d^5$    \\  
 W       & 5$p^6$ 6$s^1$ 5$d^5$    \\  
 Cr      & 3$p^6$ 3$d^5$ 4$s^1$    \\  
 La      & 5$s^2$ 5$p^6$ 5$d^1$ 6$s^2$  \\  
 Cu      & 3$p^6$ 3$d^{10}$ 4$s^1$ \\  
 Co      & 3$s^2$ 3$p^6$ 3$d^8$ 4$s^1$           \\
 Ce      & 5$s^2$ 5$p^6$ 4$f^1$ 5$d^1$ 6$s^2$  \\
 \bottomrule
 \end{tabular*}
 \end{table}

Generalized Gradient Approximation (GGA-PBE) was used here for the exchange-correlation functional~\cite{1996PER}.
For visualization of atomic structures and charge densities, VESTA was used~\cite{2011MOM}.
To compute effective masses of the charge carriers, ``effmass" python package was used~\cite{2018WHA}.

To model doped anatase systems, a single Ti atom in the 2 $\times$ 2 $\times$ 2 (96 atom supercell built from the conventional unit cell of anatase) was substituted by the dopant atom. 
The doping concentration thus achieved was 1.042 at\% (defined as the percent of total atoms substituted by the dopant) or 3.125 at\% (defined as the percent of Ti atoms substituted by the dopant).
500 eV was used as the cutoff for plane wave expansion.
A uniform \textit{k} spacing of 0.20 {\AA}$^{-1}$ along all the three reciprocal axes was used to generate the \textit{k} point grid.
Convergence tests were carried out to determine these parameters.
SCF total energies were converged within $10^{-6} eV$.
All the doped structures were first fully relaxed (lattice parameters + atomic position optimization) using the conjugate-gradient (CG) algorithm.
The relaxation was stopped when the norms of all the forces were smaller than 0.001 eV{\AA}$^{-1}$.

\subsection{Determination of Hubbard $U$ from first principles}
\label{u_determination}
To compute the Hubbard $U$ parameter from first principles, the linear response method developed by Cococcioni \textit{et al.} was utilized~\cite{2005COC}.
We briefly describe the method here.
For a rigorous derivation and other finer details, we refer interested readers to the original paper~\cite{2005COC}.
The basic idea is to apply a potential to the correlated ($d$ or $f$ orbitals) levels of an atomic site (for example Ti) and observe how this perturbation potential changes the $d$ orbital occupancy.
The $d$ orbital electron occupancy can be computed simply from the orbital and atom decomposed charge density.
The change in the $d$ orbital occupancy with respect to the perturbation potential is quantified by two response functions: self-consistent response and non self-consistent response.
The former is computed while the charge density is allowed to relax and the latter is computed while keeping the charge density fixed.
Non self-consistent response~\cite{2005COC} can be written as:

\begin{eqnarray}\label{eq:response_nscf}
    \chi^0_{J} = \frac{\partial N_J^{NSCF}}{\partial V_J},
\end{eqnarray}

\noindent
where, ${\partial V_J}$ is the perturbation potential applied at a site $J$ and ${\partial N_J^{NSCF}}$ is the change in $d$ orbital occupations at that site.
Similarly, the self-consistent response~\cite{2005COC} can be written as:

\begin{eqnarray}\label{eq:response_scf}
    \chi^1_{J} = \frac{\partial N_J^{SCF}}{\partial V_J}
\end{eqnarray}

\noindent
Now, Hubbard parameter is given by~\cite{2005COC}:

\begin{eqnarray} \label{eq:hubbard}
    U_{eff} = {\left({\frac{\partial N_J^{SCF}}{\partial V_J}}\right)}^{-1} - {\left({\frac{\partial N_J^{NSCF}}{\partial V_J}}\right)}^{-1}
\end{eqnarray}

\begin{figure}[htbp]
	\centering
		\includegraphics[scale=.60]{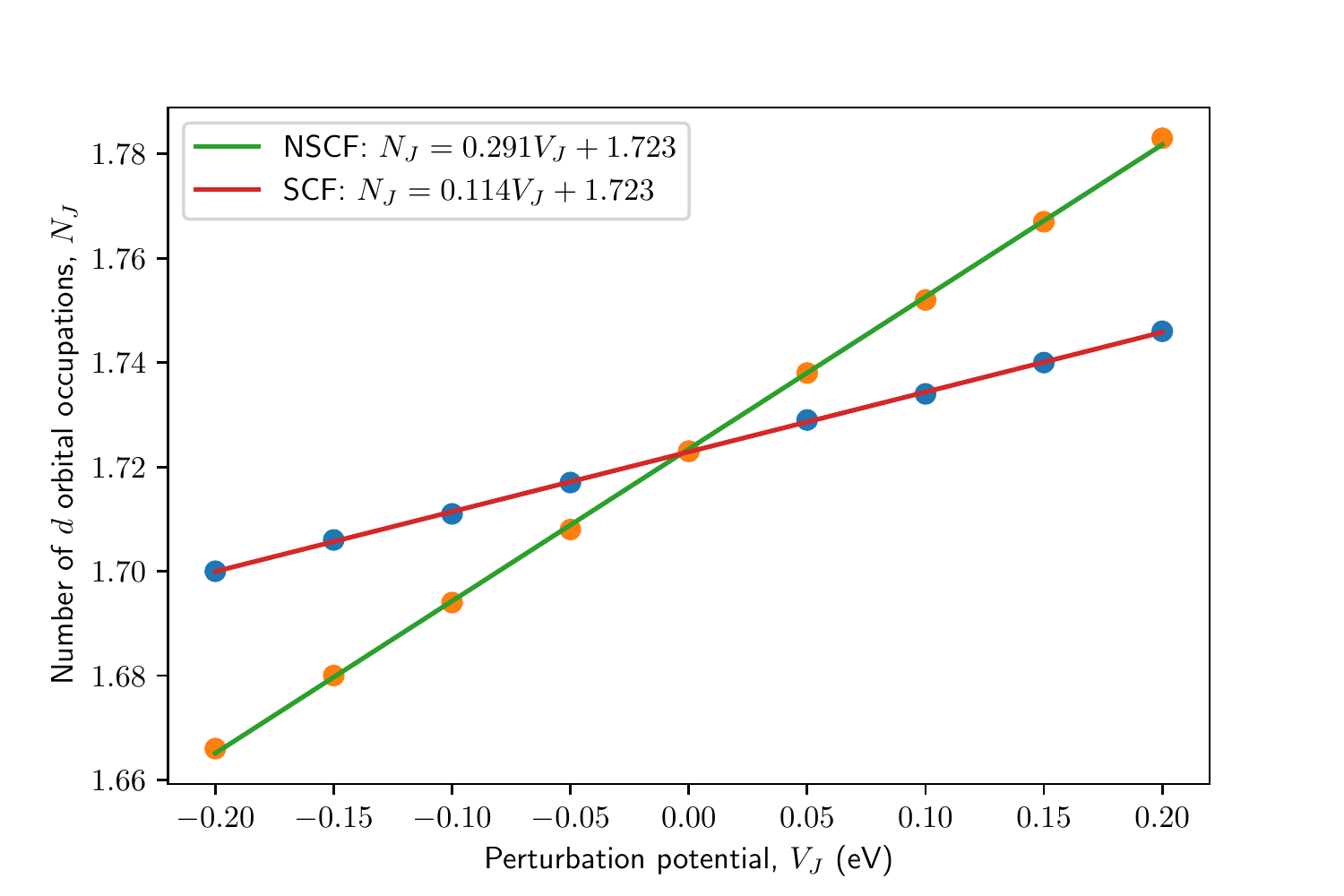}
    \caption{Linear fitting of $d$ orbital occupation ($N_J$) as a function of the perturbation potential ($V_J$) for a Ti site ($J$) in TiO$_2$. The slope of the fitted lines gives the relevant response functions; $\frac{\partial N_J^{SCF}}{\partial V_J} = 0.114$, $\frac{\partial N_J^{NSCF}}{\partial V_J} = 0.291$}
	\label{FIG:1}
\end{figure}

\noindent
In practice, the $d$ orbital occupations ($N_J$) are computed for several values of perturbation potential ($V_J$) (see Fig.~\ref{FIG:1}), followed by linear fitting to express $N_J$ as a function of $V_J$. 
The slope of this fit ($\frac{\partial N_J}{\partial V_J}$) then gives the relevant response functions (see Eq.~\ref{eq:response_nscf} and Eq.~\ref{eq:response_scf}).
Fig.~\ref{FIG:1} shows the fitting process for Ti.
$U$ for Ti is computed as follows:

\begin{eqnarray} 
    \label{eq:hubbard_ti}
    U_{Ti} = \frac{1}{0.114} - \frac{1}{0.291} = 5.29 eV
\end{eqnarray}
The computed $U$ values for the dopants are listed in the Supplementary data.

\begin{figure}[htbp]
	\centering
		\includegraphics[width=\linewidth]{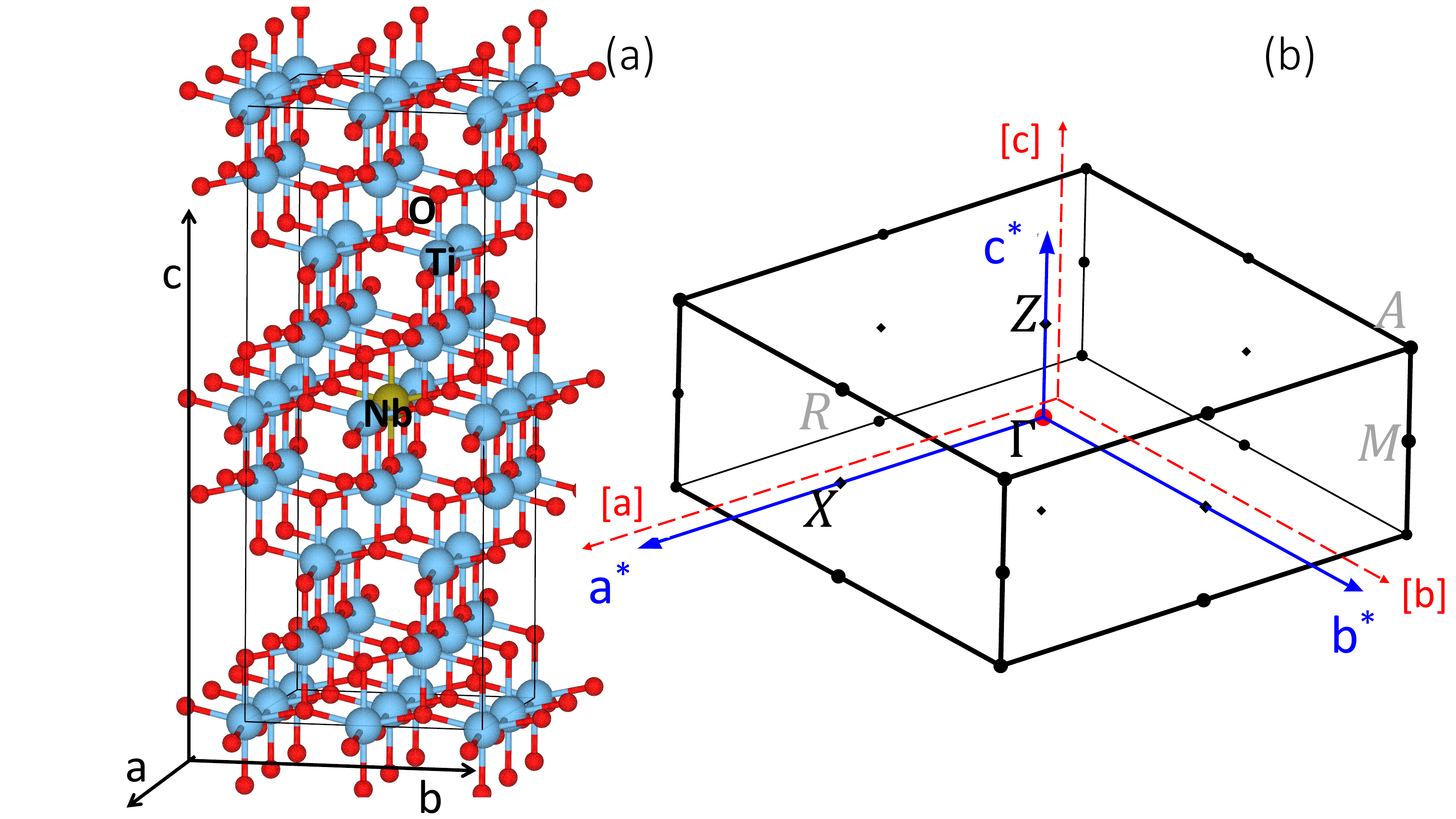}
    \caption{(a) Schematic of Nb-doped anatase tetragonal supercell (with blue Ti atoms, red O atoms and yellow Nb atom). The three primary lattice directions are shown. $c$ is along the tetragonal axis. (b) First brillouin zone and the high symmetry points for the tetragonal supercell. $a^*$, $b^*$ and $c^*$ represent the reciprocal lattice vectors and point in the same direction as real space lattice vectors $a$, $b$ and $c$ respectively.}
	\label{FIG:2}
\end{figure}

\subsection{Effective mass computation}
\label{emass_method}
Effective mass can be computed in several different ways from the band dispersion $E(k)$, obtained from a DFT calculation.
Fig.~\ref{FIG:2}(b) shows the high symmetry points in the reciprocal space.
The $k$-paths $\Gamma-X$ and $\Gamma-Z$ lie along the reciprocal lattice vectors $a^*$ (or the crystal's $a$ axis) and $c^*$ (or the crystal's orthogonal $c$ axis) respectively and hence are important for effective mass analysis.
Effective mass analysis was done on band segments in the energy range of 0.25 eV from the extrema (for example the $\Gamma$ point along the $\Gamma-X$ direction on the conduction band). 
The band segments chosen for effective mass analysis in anatase are shown in the Supplementary data (Fig.1). 
The energy range of 0.25 eV was chosen because for methods like the weighted least squares effective mass, the weights computed using the Fermi-Dirac distribution become negligible for higher energies~\cite{2019WHA}.
For maintaining consistency, the same energy range (band segment) was used for other methods too.
It is important to mention here that, the energy range (and hence the band segment) could be smaller for cases where the bands are extremely flat (for example mid-gap states where the bands are spread over energy ranges much smaller than 0.25 eV).
For methods like the finite difference and the least squares effective mass, the energy range does not mean much.
By definition, these methods only need three points and five points respectively (in the vicinity of band extrema) on the band dispersion curve.
Hence, a more important parameter would be the size of the $k$-spacing (distance in the reciprocal space between consecutive $k$-points) used for computing the band dispersion.
A $k$-spacing of $\approx$0.005{\AA}$^{-1}$ (determined by testing the convergence of finite difference effective mass w.r.t. $k$-spacing) was used consistently for all the methods.

Simpler methods like the curvature effective mass (using finite difference method) are used to compute effective mass at band edges (or extrema, for example in the vicinity of $\Gamma$ point along the $\Gamma-X$ direction on the conduction band) only.
If the bands were perfectly parabolic, the curvature at every point would be the same.
In real materials, however, the dependence of electron energy ($E$) on wave vector ($k$) can't be parabolic within the entire first Brillouin zone.
The bands are close to parabolic near the band edges only (for example in the vicinity of the $\Gamma$ point along the $\Gamma-X$ direction).
To account for this non-parabolicity, several newer definitions~\cite{2018WHA,2019WHA} for the effective mass were utilized.
The following definitions of effective mass were used:
\subsubsection{Curvature effective mass}
\label{curv_efm}
This is the conventional definition of effective mass~\cite{1976ASH}.
        \begin{eqnarray} \label{eq:curveefm}
            \frac{1}{m_c} = \frac{1}{\hbar^2} \frac{\partial^2E}{\partial k^2},
        \end{eqnarray}

        \noindent
        where $m_c$ is the curvature effective mass.
        The term $\frac{\partial^2E}{\partial k^2}$ indicates the curvature of the bands and can be computed from the band dispersion or band structure from DFT calculation.
        For parabolic bands, the dispersion relation is given by:

         \begin{eqnarray}\label{eq:curveefm1}
            E(k) = \frac{\hbar^2k^2}{2m_c}
         \end{eqnarray}

         \noindent
        This curvature terms can be algebraically computed in several ways~\cite{2018WHA, 2019WHA}:
\par
\textbf{Finite difference effective mass}: A three point finite difference equation was used to calculate the curvature of the band at point i.
                \begin{eqnarray} \label{eq:finitediff}
                    \frac{\partial^2 E}{\partial k^2} = \frac{E_{i+2} - 2E_{i+1} + E_i}{|k_{i+1} - k_i|},
                \end{eqnarray}

                \noindent
                 where $k_i$ is a point in reciprocal space and $E_i$ is the energy eigenvalue at that point.
                 $E_{i+1}$ is the next eigenvalue and so on.
                 The finite difference effective mass was used to compute effective mass value at the band edge (or extrema; for example at $\Gamma$ point along the $\Gamma-X$ direction).
\par
\textbf{Unweighted least squares fitting}: The parabolic dispersion is given by the formula:

                 \begin{eqnarray} \label{eq:leastsq}
                     E = ck^2
                 \end{eqnarray}

                 \noindent
                 To obtain the coefficient $c$, the dispersion was fitted using least-squares method by minimizing the residuals:

                  \begin{eqnarray}\label{eq:residuals}
                        \sum_{i=1}^{5} {(ck_i^2 - E_i)}^2
                  \end{eqnarray}

                  \noindent
                  Five points from band dispersion near the band edge were used for fitting.
\par
\textbf{Weighted least-squares fitting}: The following sum of squared residuals was minimized:
                  \begin{eqnarray}\label{eq:wtleastsq}
                        \sum_{i=1}^{n} W_i{(ck_i^2 - E_i)}^2
                  \end{eqnarray}

                  \noindent
                   Their weights were calculated from the Fermi-Dirac distribution:

                  \begin{eqnarray} \label{eq:f-dweight}
                    W_i(E_i, T) = \frac{1}{\exp \left(\frac{E_i-E_F}{k_BT}\right) + 1},
                  \end{eqnarray}                 
                   The summation was done over an energy range of 0.25 eV.
                   All the points in the band dispersion within this energy range were considered.
                   Due to the exponential term in the denominator in Eq.~\ref{eq:f-dweight}, points on the dispersion curves with an energy difference larger than 0.25 eV ($= 10k_BT$, where T is 300 K), have negligible weights and hence are left out from the weighed sum~\cite{2019WHA}.

\subsubsection{Transport effective mass}
\label{trans_efm}
At high temperature or at higher carrier concentration, eigenstates far from the band extrema (non-parabolic nature) are accessed.
        To account for non-parabolic nature of bands, wave-particle duality for an electron wavepacket can be used to derive the following formula for transport effective mass~\cite{2012ARI, 2019WHA}.

        \begin{eqnarray} \label{eq:transportefm}
            \frac{1}{m_t} = \frac{1}{\hbar^2 k} \frac{\partial E}{\partial k}
        \end{eqnarray}

\subsubsection{Kane quasi-linear dispersion}
To account for non-parabolic bands, another approach is to include non-linear terms and expand Eq.~\ref{eq:curveefm1} as follows:

          \begin{eqnarray}\label{eq:expansion}
                \frac{\hbar^2 k^2}{2m_{t,edge}} = E + \alpha E^2 + \beta E^3 + ...,
          \end{eqnarray}

          \noindent
          where $m_{t,edge}$ is the Kane mass at band edge.
          If degree 3 and higher terms are neglected, Kane quasi-linear dispersion relation can be obtained~\cite{1957KAN}.

           \begin{eqnarray} \label{eq:kane_dispersion}
                \frac{\hbar^2 k^2}{2m_{t,edge}} = E(1 + \alpha E),
           \end{eqnarray}

           \noindent
           where the $\alpha$ parameter specifies the non-parabolic nature of bands.
           If the bands are perfectly parabolic, $\alpha = 0$.
           For conduction bands in general $\alpha$ is positive, and it is negative for the holes in valence bands.
           Differentiating the kane dispersion w.r.t $k$ gives the transport effective mass as~\cite{2019WHA}:

           \begin{eqnarray} \label{eq:transport_mass}
               m_t(E) = m_{t, edge}(1 + 2\alpha E)
            \end{eqnarray}

            To compute $\alpha$, first the transport effective mass ($m_t(E)$, at various energy eigenvalues, $E$) was computed using Eq.~\ref{eq:transportefm} from the band dispersion.
            Eq.~\ref{eq:transport_mass} was then used to compute $\alpha$.
            
\subsubsection{Optical effective mass}
\label{opt_efm}
Optical effective mass is defined as~\cite{2019WHA}:

             \begin{eqnarray}\label{eq:optefm}
                \frac{1}{m_{opt}} = \frac{2}{n_e} \sum_{l} \sum_{k}^{occ.} \frac{1}{m_c^l(k)},
             \end{eqnarray}

             \noindent

              where $m_{opt}$ is the optical effective mass, $n_e$ is the charge carrier concentration and $m_c^l$ is the curvature effective mass for a band $l$ and an occupied eigenstate $k$.
              Optical effective mass incorporates summation over all occupied eigenstates and each band $l$.
              This can account for non-parabolicity in bands.
              According to a derivation by Huy \textit{et al.}~\cite{2011HUY}, the summation can be replaced by an integration along one-dimensional paths in the $k$-space:

               \begin{eqnarray}\label{eq:optefm2}
                    \frac{1}{m_{opt}} = \frac{\sum_{l} \int f(E, T) \frac{\partial^2 E}{\partial k^2}dk}{\sum_{l} \int f(E, T)dk},
               \end{eqnarray}

               \noindent
               where $f(E, T)$ represents the Fermi-Dirac distribution.

                Since band dispersions are rarely parabolic for real materials, Kane dispersion is more accurate.
                The computed transport and optical effective masses depend upon the charge carrier energy.
                When through doping, or increasing temperature etc. the bands are progressively filled the effective mass also changes (generally increases~\cite{2002RIF}).
                This also changes properties like charge carrier mobility.
                The amount of change in optical and transport effective masses with respect to carrier concentration depends upon non-parabolicity of bands.
                It should be noted here that, the curvature effective mass does not depend upon carrier concentration, rather is defined by the band curvatures alone.
                Since optical effective mass incorporates band non-parabolicity it has been used as an important parameter for the design of transparent conducting oxides~\cite{2013HAU}.

\section{Results \& Discussion}

\subsection{Pristine anatase}

The computed lattice parameters, bond lengths, band gap and bader atomic charges are shown in Table~\ref{tab:tabl2}.
Computations were performed for two values of $U$ parameter, 4.2 eV and 5.29 eV.
The former value was taken from the literature~\cite{2018GHA}.
The latter value was determined using linear response {\it ansatz} as discussed in Sec.~\ref{u_determination}.
The computed values are compared with both theoretical and experimental values available in the literature.
The lattice parameters {\it a} and {\it b} and the Ti-O bond lengths are found to be in agreement with the previously estimated experimental and computational data.
However, there is a small discrepancy in case of the lattice parameter {\it c}.
The computed lattice parameters depend slightly on the pseudopotential and the exchange-correlation functional used for the computation.
Labat {\it et al.} performed a thorough investigation of the dependence of lattice parameters of anatase on the computational methodology (pseudopotential + xc).
They discovered that the lattice parameters {\it a} and {\it b} are largely independent of the computational methodology used, and the associated error is within $\pm1\%$.
However for {\it c}, the dependence on computational methodology was noticeable.
It was found to be systematically overestimated by about 4$\%$ with respect to the experimental data ~\cite{2007LAB}. 
The current results are consistent with their findings.
The overall agreement of structural parameters with the existing computational and experimental reports indicates the reliability of the methodology used in this work in predicting the electronic defect states for anatase.

\begin{table*}[htbp,width=\textwidth]
    \caption{Lattice parameters, bond lengths (in {\AA}), band gap (in eV) and Bader charges in anatase.}
\label{tab:tabl2}
\begin{tabular}{@{}lllllllll@{}}
\toprule
 &
  Methodology &
  \multicolumn{2}{l}{\begin{tabular}[c]{@{}l@{}}Lattice\\ parameters\end{tabular}} &
  \multicolumn{2}{l}{\begin{tabular}[c]{@{}l@{}}Ti-O bond\\ lengths\end{tabular}} &
  $E_g$ &
  \multicolumn{2}{l}{\begin{tabular}[c]{@{}l@{}}Bader\\ charges\end{tabular}} \\ \midrule
 &
   &
  a=b &
  c &
  Apical &
  Equatorial &
   &
  Ti &
  O \\ \midrule
\multirow{4}{*}{\begin{tabular}[c]{@{}l@{}}Current\\ work\end{tabular}} &
  GGA+$U$ (4.2 eV) &
  3.84 &
  9.84 &
  1.99 &
  1.95 &
  2.44 &
  +2.25 &
  -1.15 \\
 &
  GGA+$U$ (5.29 eV) &
  3.87 &
  9.77 &
  2.02 &
  1.98 &
  2.72 &
  +2.41 &
  -1.21 \\
 &
  mBJ-GGA &
  -- &
  -- &
  -- &
  -- &
  2.6 &
  -- &
  -- \\
 &
  mBJ-LDA &
  -- &
  -- &
  -- &
  -- &
  2.8 &
  -- &
  -- \\ \midrule
\multirow{3}{*}{\begin{tabular}[c]{@{}l@{}}Other\\ computational\\ work\end{tabular}} &
  GGA+$U$~\cite{2016ARA} &
  3.83 &
  9.63 &
  2.00 &
  1.96 &
  2.00 &
  -- &
  -- \\
 &
  GGA+$U$~\cite{2017GER} &
  3.82 &
  9.55 &
  -- &
  1.95 &
  2.61 &
  -- &
  -- \\
 &
  \begin{tabular}[c]{@{}l@{}}Full-potential\\ all-electron\\ calculation (PBE)~\cite{2017KOC}\end{tabular} &
  3.81 &
  9.72 &
  2.01 &
  1.95 &
  -- &
  +2.50 &
  -1.30 \\ \midrule
\begin{tabular}[c]{@{}l@{}}Experimental\\ value~\cite{2008REY,1987BUR}\end{tabular} &
   &
  3.78 &
  9.50 &
  1.98 &
  1.93 &
  3.21 &
  -- &
  -- \\ \bottomrule
\end{tabular}
\end{table*}

\begin{figure}[htbp]
	\centering
		\includegraphics[width=\linewidth]{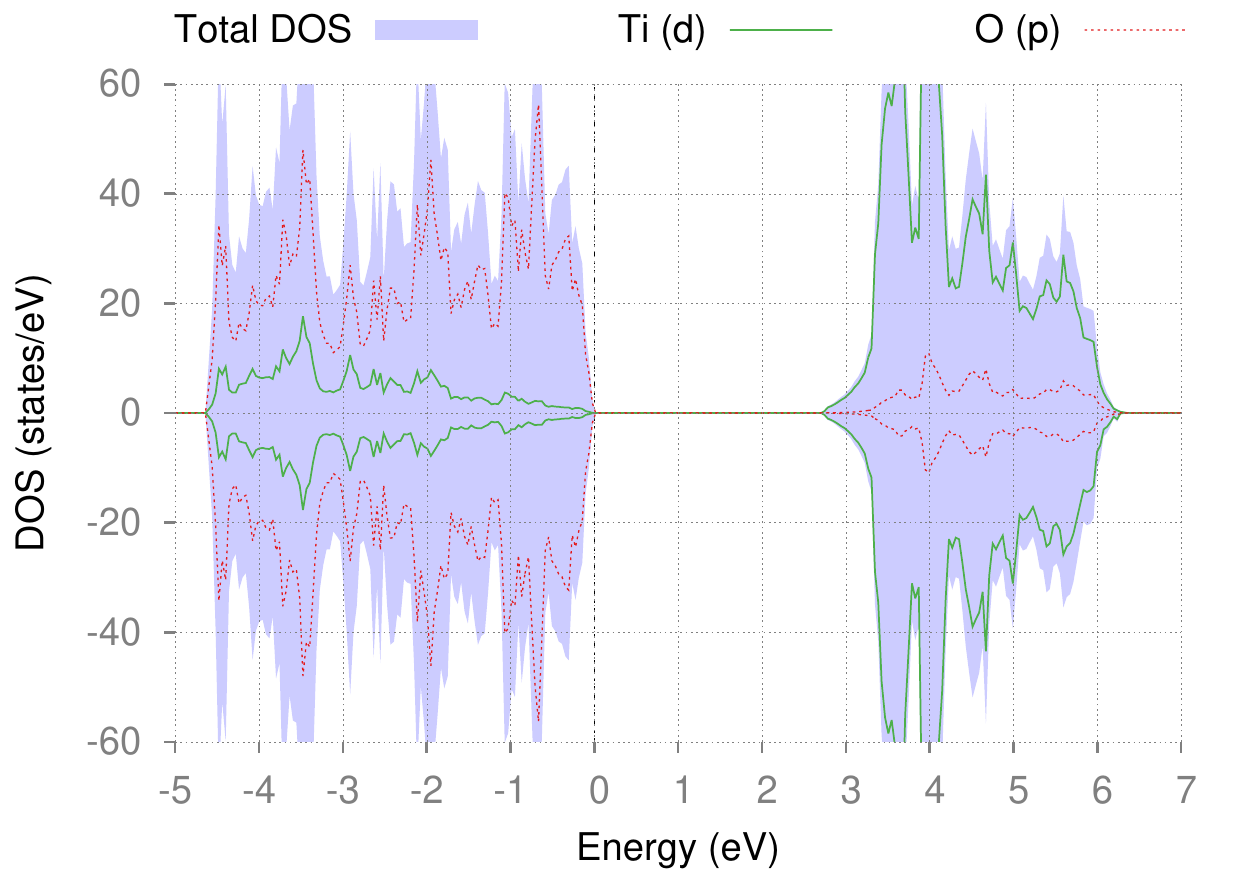}
    \caption{Projected DOS for pristine TiO$_2$ obtained using GGA+$U$ (=5.29 eV). Fermi level has been shifted to 0 eV. A band gap of 2.72 eV was obtained.}
	\label{FIG:3}
\end{figure}

The average bader charges~\cite{2011YU,2006HEN} on Ti and O atoms also agree with the previous literature reports. 
The charges computed using $U = 5.29 eV$ are found to be more accurate.
The band structure showed an indirect band gap of 2.44 eV ($U = 4.2eV$) and 2.72 eV ($U = 5.29 eV$) which is an underestimation even after Hubbard $U$ correction (see Fig.~\ref{FIG:3} for DOS plot of anatase).

Clearly, the larger $U$ value opens up the gap more and the band gap value is closer to those obtained using mBJ-GGA and mBJ-LDA.

\subsection{Electronic structure of doped anatase systems}
Projected DOS plots of various doped systems are shown in Fig.~\ref{FIG:4}, ~\ref{FIG:5} and ~\ref{FIG:6} and the band structures are shown in the Supplementary data.
The computed lattice parameters are also mentioned in the Supplementary data. 
The dopants Mo, W, Cr, Cu and Co formed mid gap states.
Mid-gap states might adversely affect for TCO applications because of the possibility of electron transition to/from the mid-gap level by absorbing certain wavelengths of light, thus reducing the transparency.
In case of Cr, Cu and Co, the mid gap states were unoccupied.
For W, the mid gap state was completely occupied.
In case of Mo, the mid gap states were partially occupied.
The mid gap states in these doped systems, were associated with localized $d$ orbital states on the dopant atoms.

\begin{figure}[htbp]
	\centering
		\includegraphics[width=\linewidth]{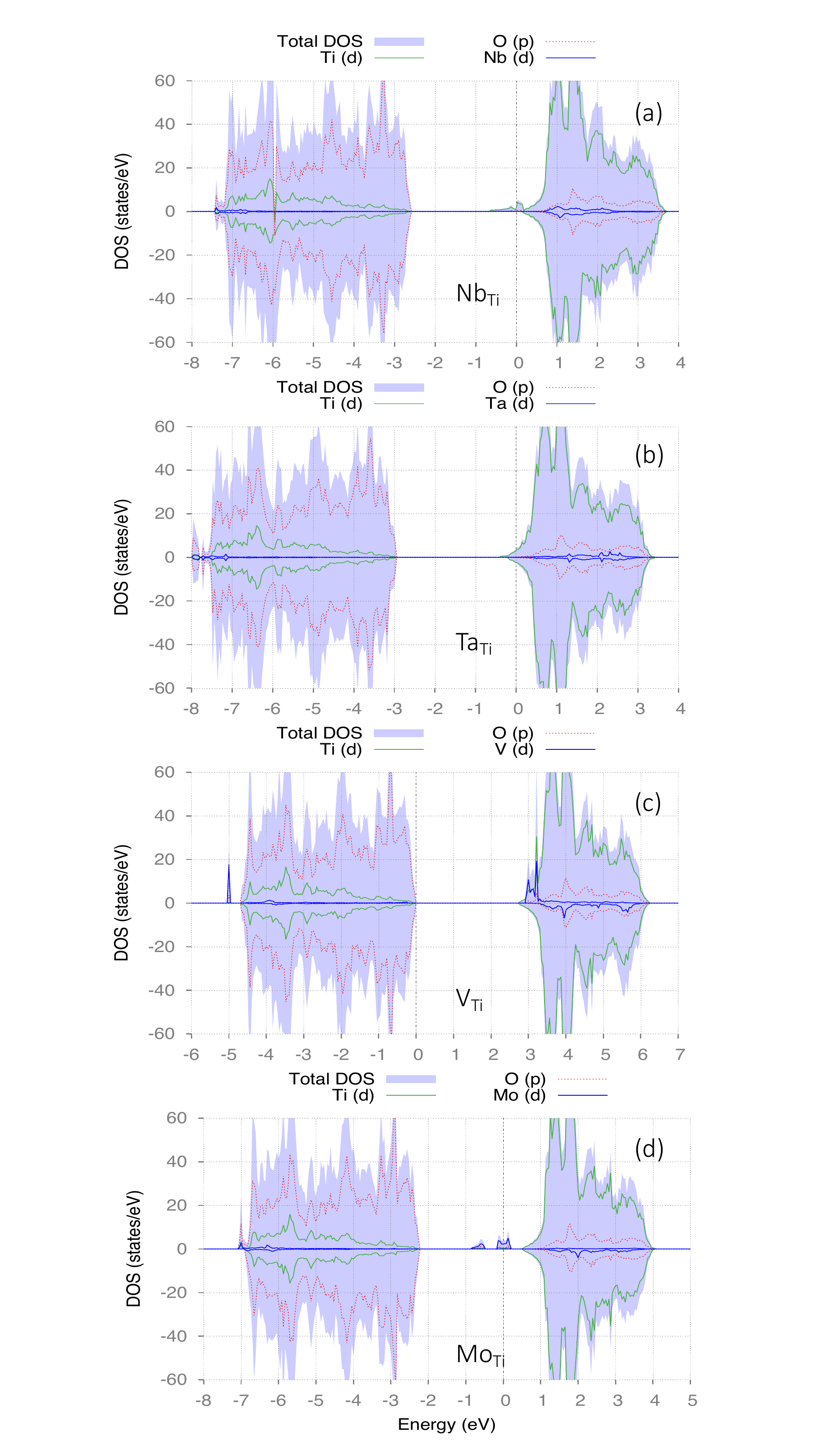}
    \caption{Projected DOS for (a) Nb-, (b) Ta-, (c) V- and (d) Mo-doped TiO$_2$. Fermi energy has been shifted to 0 eV.}
	\label{FIG:4}
\end{figure}

\begin{figure}[htbp]
	\centering
		\includegraphics[width=\linewidth]{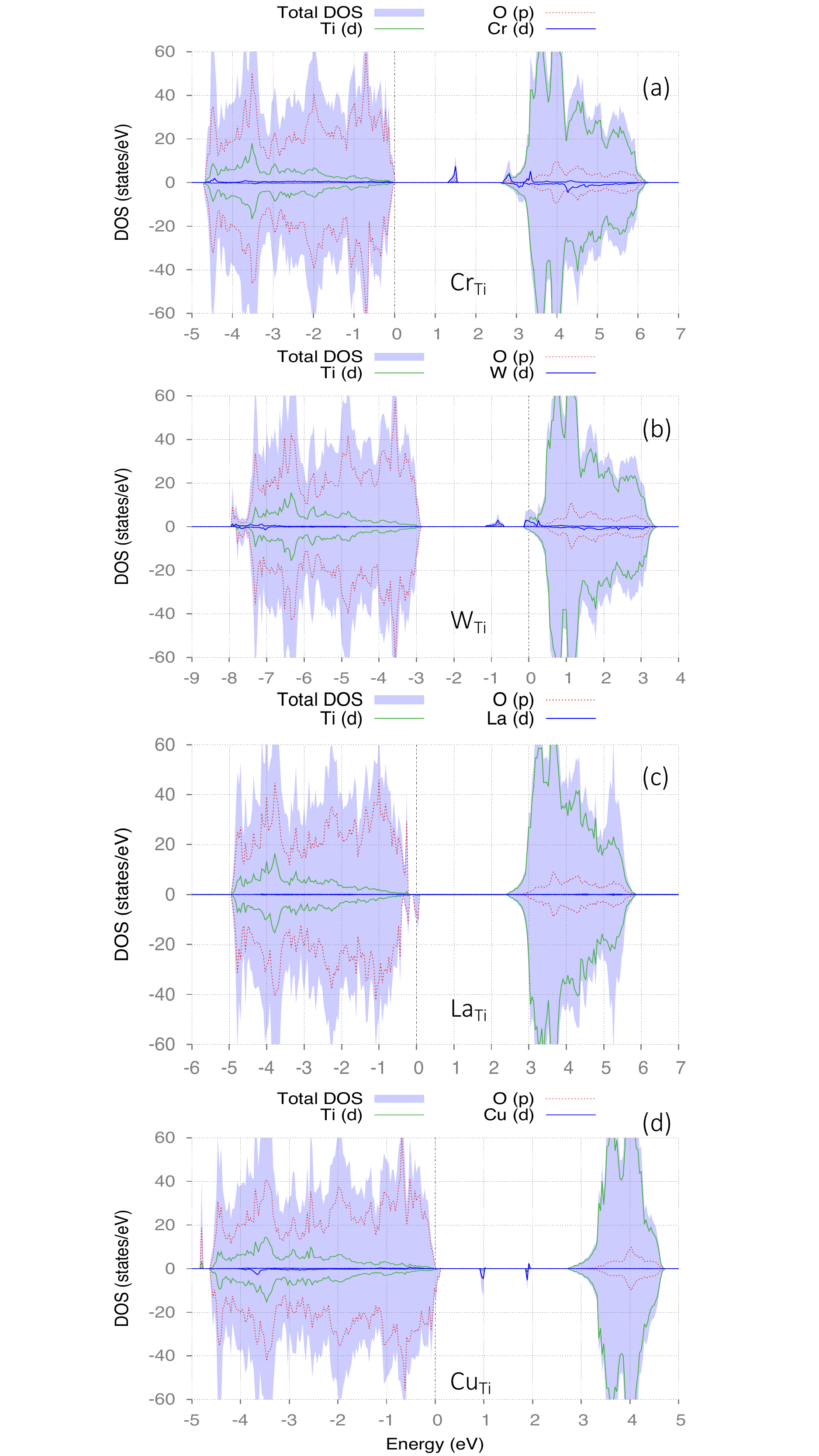}
    \caption{Projected DOS for (a) Cr-, (b) W-, (c) La- and (d) Cu-doped TiO$_2$. Fermi energy has been shifted to 0 eV.}
	\label{FIG:5}
\end{figure}

Fig.~\ref{FIG:7}(a)-(c) show the charge density isosurfaces plotted at the mid gap states for Mo, W and Cr dopants respectively. 
All these mid gap states are clearly localized on the dopant and a few neighbouring atoms.
In case of Mo, the mid gap state was associated with the $d$ states on the dopant Mo atom, and $p$ states on the six neighbouring O atoms.
For Cr-doped anatase, the mid gap states were localized on the dopant Cr atom (as $d_{xy}$ orbitals states) and four neighbouring O (as $p$ states) atoms.
For Cu and Co, two empty mid gap states were formed (referred to as: closer to the valence band edge and closer to the conduction band edge) (see Fig.~\ref{FIG:5}(d) and Fig.~\ref{FIG:6}(a) respectively).
Fig.~\ref{FIG:8} shows the corresponding charge isosurface plots.
In case of Cu, states closer to the valence band edge were associated with Cu $d_{z^2}$ and O $p$ (on neighbouring O atoms) states.
The ones closer to the conduction band were composed of Cu $d_{x^2-y^2}$ and O $p$.
Similarly for Co, states closer to the valence band edge were composed of Co $d_{xy}$ and O $p$ (on four neighbouring O atoms), while the ones closer to the conduction band were associated with Cu $d$ and O $p$ (on six neighbouring O atoms).

The dopant Ta formed delocalized Ta $d$ states in the conduction band and the Fermi energy was positioned in the conduction band near the conduction band edge, thus giving Ta-doped TiO$_2$ a metallic nature (see Fig.~\ref{FIG:4}(b)).
This would mean that Ta-doped TiO$_2$ would show improved electronic conductivity.
A previous DFT study also concluded the metallic nature of Ta-doped TiO$_2$~\cite{2017MUH}.
The band gap was essentially unchanged, which could mean that Ta-doped TiO$_2$ to have transparency similar to that of pristine TiO$_2$.
Improved conductivity in Ta-doped anatase thin films has also been observed experimentally~\cite{2018SUH,2015ANI,2005HIT,2018ZHA}.
Nb dopant formed delocalized Nb $d$ states (see Fig.~\ref{FIG:4}(a)) in the conduction band with no mid-gap states.
Han \textit{et al.} also found no mid gap states for Nb-doped anatase in their DFT study~\cite{2013HAN}.
However, additional Ti $d$ states were also formed at the conduction band edge.
The Fermi level was positioned near the conduction band edge and the electrons occupying the defect states at the conduction band edge could transition to the conduction band.
This could lead to improved \textit{n}-type conductivity in Nb-doped TiO$_2$.
However, there was a reduction in the band gap by 0.78 eV, which could reduce the transparency of Nb-doped TiO$_2$ films.
Improved conductivity in Nb-doped anatase has also been observed experimentally~\cite{2015ANI,2005FUR,2016LU,2015ZHA, 2021MAN}.
Similar electronic structures (DOS plots) for Nb and Ta were also obtained in previous DFT studies~\cite{2017MUH}.
In case of W-doped anatase, W $d$ orbital associated states were formed both in the mid-gap and also at the conduction band edge (see Fig.~\ref{FIG:5}(b)).
The Fermi level was positioned in the states formed at the conduction band edge and the mid-gap state was fully occupied.
The position of fermi level suggests that electrons are injected in the conduction band which could provide \textit{n}-type conductivity to W-doped anatase.
Unlike other dopants forming mid-gap states, W dopant formed only a single occupied mid-gap state.
This could reduce transparency but less when compared to other dopants forming multiple mid-gap states.
Multiple mid-gap states would lead to many possible electron transitions from/to the mid gap states which could significantly reduce transparency. 
The charge density isosurface plotted for the mid gap state in W-doped TiO$_2$ is shown in Fig.~\ref{FIG:7}(b).
The mid gap state was found to be mainly composed of localized states on the W atom ($d_{xy}$), and four nearby O ($p$) atoms.

\begin{figure}[htbp]
	\centering
		\includegraphics[width=\linewidth]{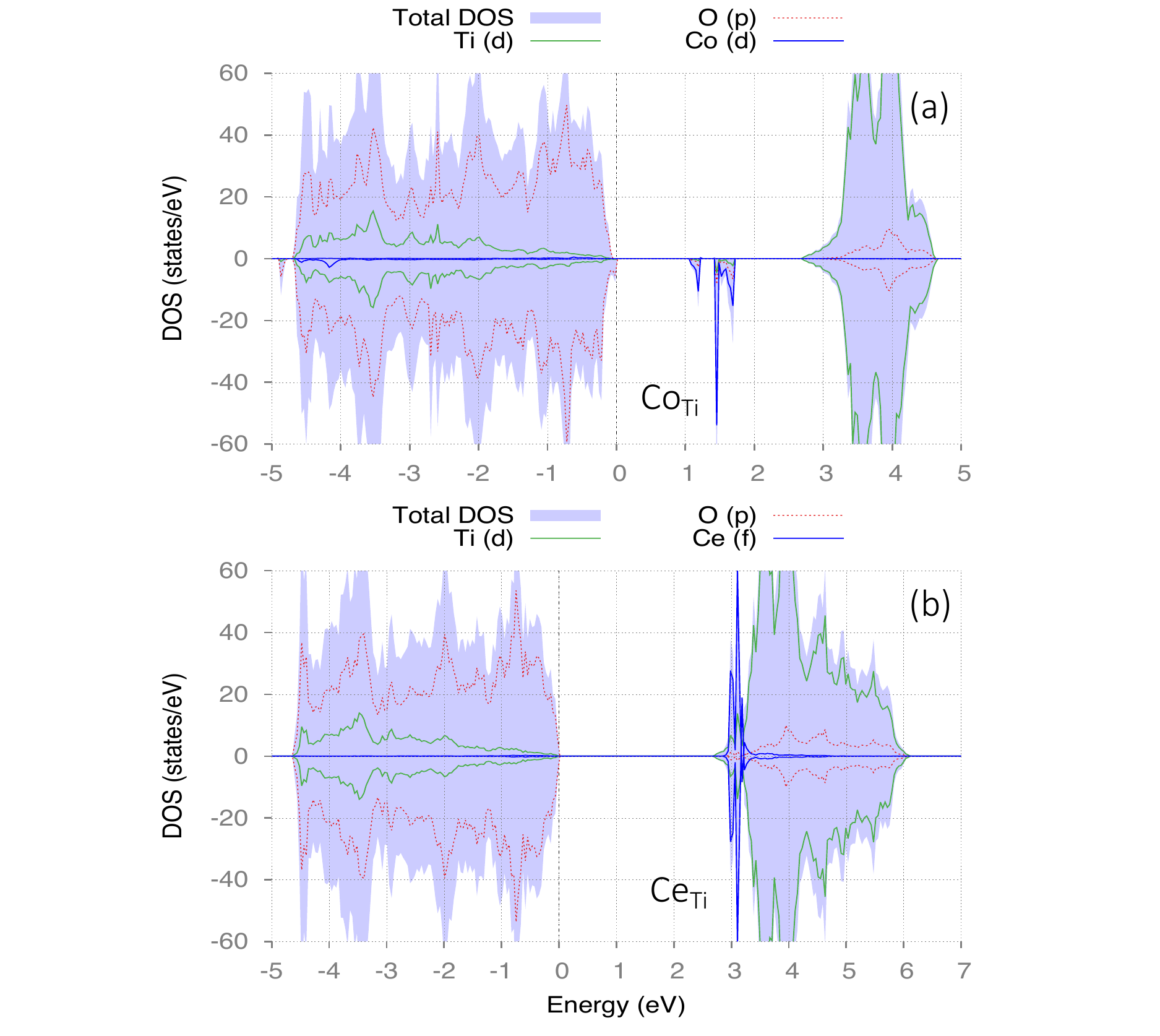}
    \caption{Projected DOS for (a) Co- and (b) Ce-doped TiO$_2$. Fermi energy has been shifted to 0 eV.}
	\label{FIG:6}
\end{figure}

\begin{figure}[htbp]
	\centering
		\includegraphics[width=\linewidth]{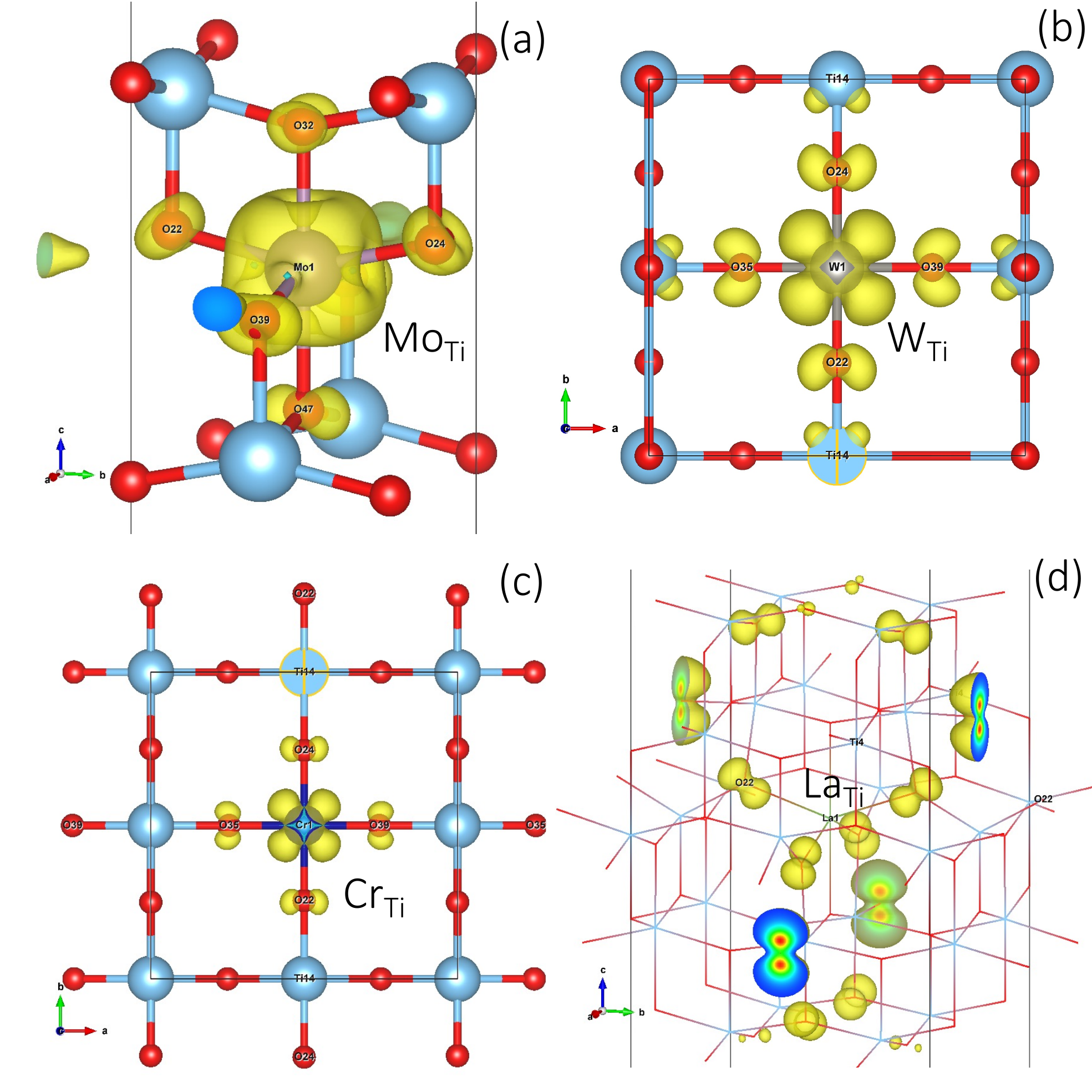}
    \caption{Charge density isosurfaces at the mid gap states plotted for various doped systems; (a) Mo-doped, (b) W-doped (plotted on a layer containing the dopant atom) and (c) Cr-doped (plotted on a layer containing the dopant atom). (d) La-doped plotted for the states at the conduction band edge. The larger blue atoms are the Ti atoms and smaller red ones are O atoms. The dopant atom substitutes the Ti atom at the center.}
	\label{FIG:7}
\end{figure}

\begin{figure}[htbp]
	\centering
		\includegraphics[width=\linewidth]{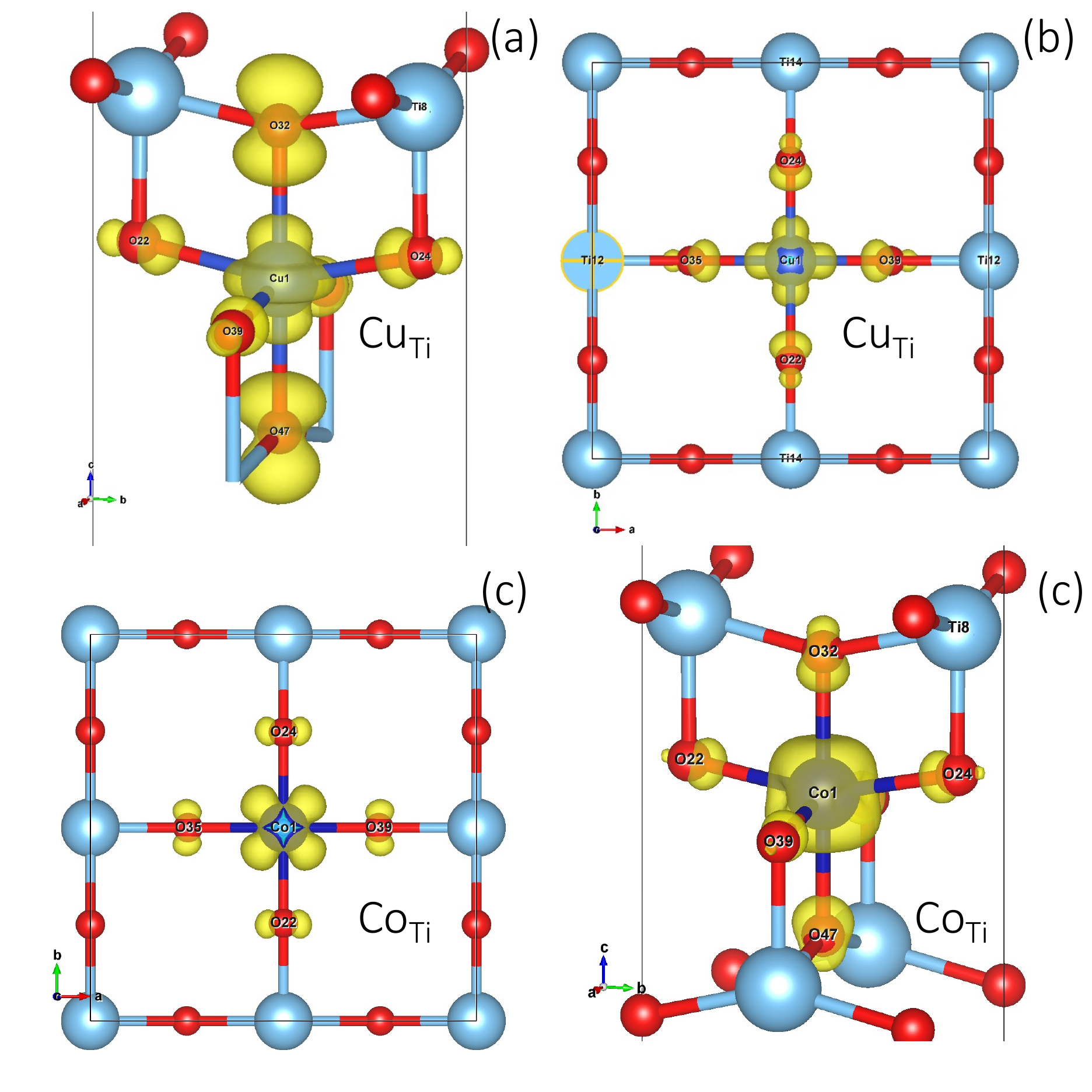}
    \caption{Charge density isosurfaces at the mid gap states plotted for various doped systems; (a) Cu-doped (for state closer to the valence band), (b) Cu-doped (for state closer to the conduction band) and (c) Co-doped (for state closer to the valence band), (d) Co-doped (for state closer to the conduction band). The larger blue atoms are the Ti atoms and smaller red ones are O atoms. The dopant atom substitutes the Ti atom at the center.}
	\label{FIG:8}
\end{figure}

Dopant V atom, formed localized $d$ states at the conduction band edge and delocalized $d$ states in the conduction band (see Fig.~\ref{FIG:4}(c)).
The Fermi level was positioned at the valence band edge.
In case of La dopant, Ti $d$ type states were formed near the valence band edge and the Fermi energy was positioned near the valence band edge (see Fig.~\ref{FIG:5}(c)).
There was a band gap reduction of 0.38 eV due to states forming at the valence band edge.
Fig.~\ref{FIG:7}(d) shows that the states formed were associated with $p$ type states on several O atoms.
Ce dopant, formed localized $f$ orbital states at the conduction band edge with the Fermi level lying at the valence band edge (see Fig~\ref{FIG:6}(b)).
There was only a slight reduction ($\approx$ 0.04 eV) in the band gap.
Zhao \textit{et al.} studied the effect of lanthanide doping~\cite{2008ZHA} on the electronic structure of TiO$_2$.
For both La and Ce dopant, they observed a reduction in the band gap (more in the case of La) which is also consistent with the current work.
Moreover, for Ce dopant they found delocalized Ce $f$ being formed in the conduction band.
In the current work however, Ce dopant was found to form localized $f$ states near the conduction band edge.
This discrepancy could be attributed to the over-delocalization of the $f$ states due to the well-known self-interaction error in DFT (with conventional XC functionals like the LDA, used in the previous study)~\cite{2011SHO}.
The current results for Ce-doped anatase however are in good agreement with the study done by Chen \textit{et al.}, which used the Hubbard $U$ correction for the highly correlated $d$ and $f$ electrons~\cite{2012CHE}.

\subsection{Effective mass analysis}

\begin{figure}[htbp]
	\centering
		\includegraphics[width=0.6\linewidth]{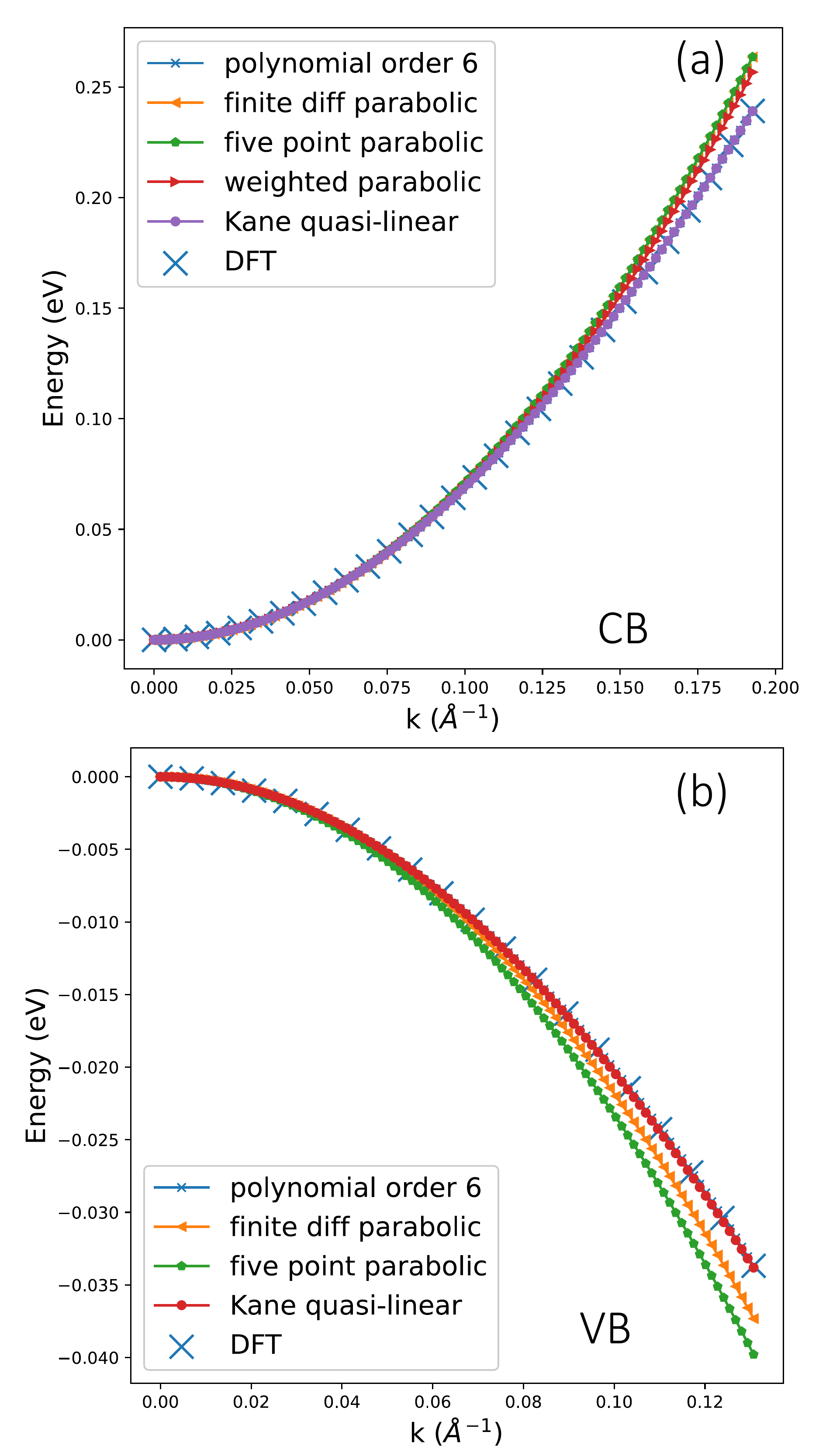}
	\caption{Band dispersion fitting at the conduction band edge (a) and valence band edge (b), along $\Gamma-X$ direction.
    A band segment with an energy range of 0.25 eV was used for the dispersion fitting of the conduction band.
    Due to the flatter nature of the valence band, the band segment in this case had a much smaller energy range. 
    The choice of band segments is shown in the Supplementary data (Fig. 1).}
	\label{FIG:9}
\end{figure}

The finite difference effective masses (in terms of $m_e$) for electrons at the conduction band edge along the $\Gamma-X$ and the $\Gamma-Z$ directions in the reciprocal space were computed to be 0.54 and 5.10 respectively. 
Clearly, there is a large anisotropy in the effective masses (and perhaps the mobility) of electrons.
Bands are much flatter along $\Gamma-Z$ direction, than the $\Gamma-X$ direction.
This anisotropy is found to be consistent with previous experimental and computational results~\cite{2008HIT,2009KAM,2011HUY}.
The anisotropy becomes quite important, when anatase films are fabricated for optoelectronic applications, because the anisotropy could manifest as different electronic conductivities along different directions in anatase films.
$\Gamma-X$ is the dominant direction for electronic conductivity, because of the lower electronic effective mass.
Due to this reason, effective mass analysis for doped systems has only been done along the $\Gamma-X$ direction in the current work.
The finite difference effective mass of holes at the valence band edge along $\Gamma-X$ direction was found to be 1.74, which is more than 3 times higher than that of the electrons at the conduction band edge. 
This could mean that holes along valence band edge are less mobile than electrons at the conduction band edge. 
This also shows that doping anatase \textit{p}-type with acceptor defects might not be very beneficial in increasing conductivity because of the higher effective mass (and low mobility) of holes in the valence band. 
The calculated values of effective masses for electrons and holes in pristine anatase along the $\Gamma-X$ direction are summarized in Table.~\ref{tab:tabl3}. 
Fig.~\ref{FIG:9} shows the fitting of DFT band dispersion for the conduction and valence bands respectively.

\begin{table}[htbp,width=\linewidth]
    \caption{Charge carrier effective masses at the band edge (extrema), along $\Gamma-X$ direction in pristine anatase}
\label{tab:tabl3}
\begin{tabular}{@{}lcccc@{}}
\toprule
         & \multicolumn{3}{l}{Effective masses (in terms of $m_e$)} & \multirow{2}{*}{$\alpha$} \\ \cmidrule(r){1-4}
         & Finite diff.        & Least sq.        & Kane mass       &                           \\ \cmidrule(l){1-5} 
Electron & 0.54                & 0.54             & 0.54            & 0.44                      \\
Hole     & 1.74                & 1.64             & 1.79            & -2.21                     \\ \bottomrule
\end{tabular}
\end{table}

$\alpha$ quantifies the non-parabolicity of the bands. 
All the computed values for the electron effective mass were found to be approximately equal with the absolute value of $\alpha$ being quite low. 
The optical effective mass for electron was computed to be 0.55, which is again almost equal to the finite difference effective mass.
This means that the conduction band considered here was highly parabolic. 
For hole electron masses, however the effective mass values varied and the absolute value of $\alpha$ was found to be higher. 
Hence, it can be concluded that the non-parabolic nature is more in the case of valence band edge than that of the conduction band edge.

\begin{figure}[htbp]
	\centering
		\includegraphics[width=\linewidth]{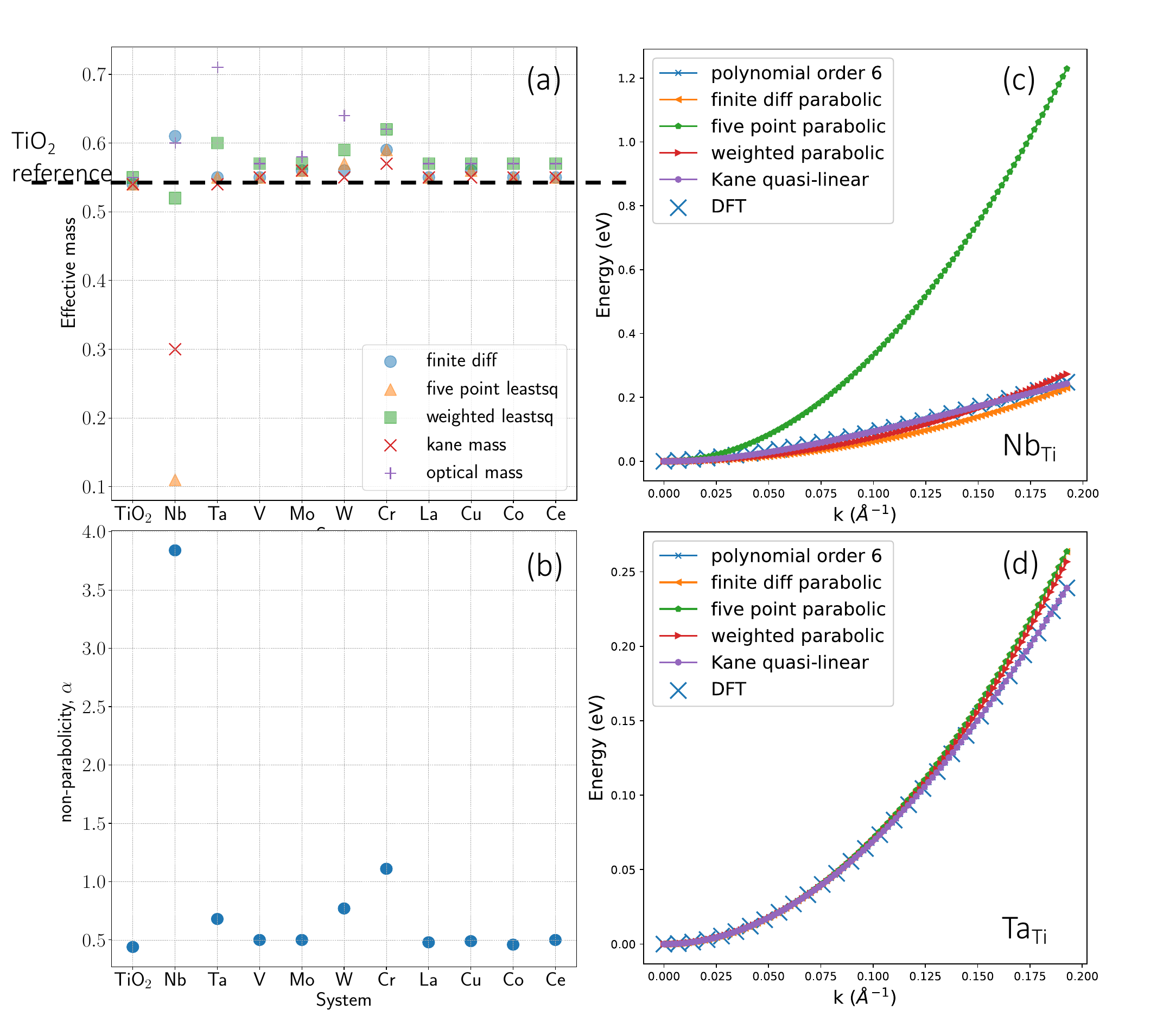}
    \caption{(a) Computed effective masses of electrons for various doped systems and (b) corresponding $\alpha$ values. 
    Band dispersion fitting at the conduction band edge along $\Gamma-X$ direction for (c) Nb-doped anatase and (d) Ta-doped anatase.
    A band segment with an energy range of 0.25 eV was used for band dispersion fitting (In case of Nb$_{Ti}$, five point parabolic fitting has been extrapolated beyond 0.25 eV, to keep the $k$ range same for all types of fitting). }
	\label{FIG:10}
\end{figure}

Fig.~\ref{FIG:10}(a) depicts computed effective mass values for various doped systems and Fig.~\ref{FIG:10}(b) shows the non-parabolicity parameter $\alpha$. 
The band dispersion was found to become more non-parabolic when dopants were introduced ($\alpha$ for all the doped systems was found to be higher than pristine TiO$_2$). 
The computed effective mass values (using various definitions) for Nb-, Ta-, W-, and Cr-doped anatase were more separated from each other than the other dopants. 
The disagreement was maximum in the case of Nb dopant. 
Interestingly, the non-parabolicity ($\alpha$) parameter for these dopants was also higher as compared to other dopants. 
Nb had the maximum value of $\alpha$. 
This shows that, these four systems have more non-parabolicity in bands (at conduction band edge) than other systems. 
The non-parabolicity of Nb- and Ta-doped anatase can also be seen when fitting of the band dispersion is attempted (see Fig.~\ref{FIG:10}(c) and (d)). 
The various fitting polynomials clearly deviate from each other (more in the case of Nb dopant). 
For non- parabolic bands, optical effective mass gives a better description because it takes into account the non-parabolicity~\cite{2019WHA}.
Optical effective masses of all the doped systems increased with respect to pristine anatase. 
This increment was more in the case of dopants Nb, Ta and W meaning that these dopant atoms perturb the conduction band edge more than other dopants. 
The values of effective masses for Nb-doped anatase computed in this work agreed with previous literature reports~\cite{2011HUY}. 
Literature reports discussing effective masses of other doped systems are rare.

\begin{figure}[htbp]
	\centering
		\includegraphics[width=\linewidth]{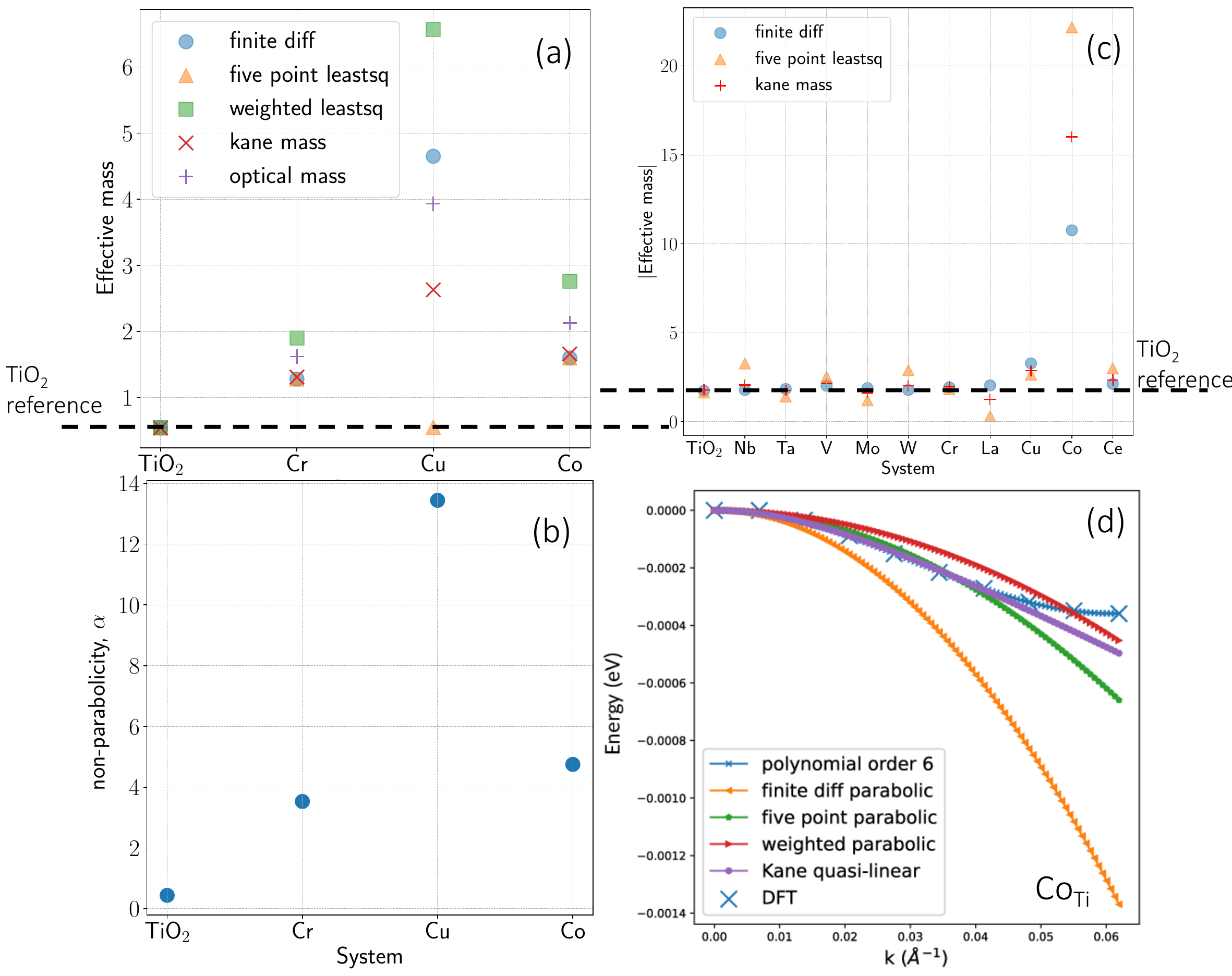}
    \caption{(a) Computed effective masses at mid-gap states for various doped systems and (b) corresponding $\alpha$ values. (c) Effective masses of holes at the valence band edge and (d) band dispersion fitting at the valence band edge along $\Gamma-X$ direction for Co-doped anatase. Valence band edge became quite flat in Co-doped anatase, which is visible from the quite small energy range on the Y-axis.}
	\label{FIG:11}
\end{figure}

Effective mass for the mid-gap states was also computed for a few cases (see Fig.~\ref{FIG:11}(a)). 
All the mid-gap states had significantly higher effective masses than the effective mass of pristine anatase at the conduction band edge.  
Moreover, $\alpha$ for the mid-gap states was also found to be significantly higher than pristine anatase (see Fig.~\ref{FIG:11}(b)), indicating high non-parabolicity. 
If optical effective masses are compared for the mid-gap states, Cu-doped anatase had the highest effective mass for mid gaps. The flat nature of these bands and higher effective masses means that electrons in mid-gap states have quite low mobility and are localized.

Effective masses were also computed for holes at the valence band edge (see Fig.~\ref{FIG:11}(c)). 
The maximum increase in effective mass was seen in the case of Co-doped anatase, which could mean that Co dopant caused the maximum perturbation of the valence band among all the dopants considered. 
It also had the most non-parabolic bands and highest value of $\alpha$ among all the dopants (see Fig.~\ref{FIG:11}(d)). The non-parabolic nature of bands led to the disagreements in effective masses computed using different algebraic definitions.

\section{Conclusion \& Summary}

In this work electronic structure calculations and effective mass analysis for doped anatase systems were performed using DFT+$U$ ($U$ computed from first principles using linear response method).
Some dopants like Mo, W, Cr, Cu and Co formed mid-gap states, others like Nb, Ta, V, La and Ce did not form any mid gap states. 
Among the dopants forming mid-gap states, Mo, Cu and Co formed multiple mid gap states. 
Such states could reduce the transparency of the host anatase material. 
In case of W, only a single occupied mid-gap state was formed, which could reduce transparency but quite less than the other dopants forming multiple mid-gap states. 
The Fermi level was positioned in the conduction band in the case of Nb, Ta and W, which would mean that these doped systems would have improved \textit{n}-type conductivity.
Thus, both conductivity and transparency may be achieved in Nb, Ta and W doped anatase.
Among these dopants, Ta would show maximum transparency. 
Transparency would be slightly reduced in case of Nb and W, due to the band gap reduction in Nb and the presence of a mid-gap state in W.

In the effective mass analysis for pristine anatase, we were able to reproduce the observed anisotropy in the effective masses of electrons along perpendicular directions.
Effective mass for electrons along the $\Gamma-Z$ direction was found to be almost 10 times higher than along the $\Gamma-X$ direction.
Dopants increased the effective masses of electrons at the CBM.
The increment in optical effective mass was higher for the dopants Ta, Nb and W than the other dopants.
It was also observed that dopants increased the non-parabolicity of bands and a parameter $\alpha$ was used to quantify this non-parabolicity.
Nb-doped system showed a significantly high value of $\alpha$, which indicates high non-parabolicity at the conduction band edge for this system.
Due to this non-parabolicity, the effective mass values (computed using various definitions) for this system differed significantly from each other.
Finally, effective masses calculated for electrons in the mid gap states were found to be quite higher ($\approx$10 times more than electrons at CBM) indicating the flat nature of mid gap states.
This could mean that the electrons in these states would be highly localized.

\section{Data availability}
The raw data required to reproduce these findings is available within the article.

\section*{Acknowledgements}
The computations in this work were performed on the supercomputer Kagayaki, provided by the Research Center for Advanced Computing Infrastructure at JAIST.
A.R. would like to gratefully acknowledge the financial support from the MEXT, Japan in the form of MEXT (Monbukagakusho) scholarship. 
K.H. is grateful for financial support from MEXT-KAKENHI (JP19K05029, JP21K03400, JP21H01998, and JP22H02170), and the Air Force Office of Scientific Research
(Award Numbers: FA2386-20-1-4036). 
R.M. is grateful for financial supports from MEXT-KAKENHI (21K03400 and 19H04692), from the Air Force Office of Scientific Research (AFOSR-AOARD/FA2386-17-1-4049;FA2386-19-1-4015), and from JSPS Bilateral Joint Projects (JPJSBP120197714). 
E.P. would like to acknowledge financial support from Science and Engineering Research Board, Department of Science and Technology, Government of India (Project No. EMR/2016/001182).

\section*{Declaration of Competing Interest}
The authors declare that they have no known competing financial interests or personal relationships that could have appeared to influence the work reported in this paper.

\appendix
\section{Appendix: Supplementary data}
Supplementary data to this article contains, computed $U$ values, lattice parameters and bond lengths of doped systems, and band structures of doped systems. 

\printcredits


\bibliographystyle{elsarticle-num}

\bibliography{manuscript.bib}





\end{document}


\begin{center}
    \LARGE \textbf{Supplementary Information}
\end{center}

\section{Computed \textit{U} values using linear response method}
\begin{table}[h]
\begin{center}
\begin{tabular}{@{}lc@{}}
\toprule
Element  & $U_{eff}$ value (eV) \\ \midrule
Ti ($d$) & 5.23                 \\
Nb ($d$) & 1.21                 \\
Ta ($d$) & 3.5                  \\
Mo ($d$) & 2.9                  \\
W ($d$)  & 4.08                 \\
Cr ($d$) & 5.78                 \\
V ($d$)  & 8.02                 \\
La ($d$) & 2.19                 \\
Cu ($d$) & 8.68                 \\
Co ($d$) & 7.05                 \\
Ce ($f$) & 6.62                 \\ \bottomrule
\end{tabular}
\end{center}
\end{table}

\section{Computed lattice parameters and bond lengths (in {\AA}) for doped systems.}
\begin{table}[h]
\begin{center}
\begin{tabular}{@{}lcccc@{}}
\toprule
                         & \multicolumn{2}{c}{Lattice parameters}                      & \multicolumn{2}{c}{Ti-O bond lengths}                       \\ \cmidrule(l){2-5} 
\multirow{-2}{*}{System} & $a=b$                        & $c$                          & Apical                       & Equatorial                   \\ \cmidrule(r){1-5}
TiO$_2$                  & 3.870                        & 9.768                        & 2.019                        & 1.981                        \\
Nb$_{Ti}$                & {\color[HTML]{3531FF} 3.877} & {\color[HTML]{3531FF} 9.775} & {\color[HTML]{3531FF} 2.054} & {\color[HTML]{FE0000} 1.966} \\
Ta$_{Ti}$                & {\color[HTML]{3531FF} 3.879} & {\color[HTML]{FE0000} 9.760} & {\color[HTML]{3531FF} 2.037} & {\color[HTML]{3531FF} 1.986} \\
V$_{Ti}$                 & {\color[HTML]{3531FF} 3.874} & {\color[HTML]{FE0000} 9.737} & {\color[HTML]{FE0000} 1.950} & {\color[HTML]{3531FF} 1.994} \\
Mo$_{Ti}$                & {\color[HTML]{3531FF} 3.877} & {\color[HTML]{FE0000} 9.759} & {\color[HTML]{3531FF} 2.042} & {\color[HTML]{3531FF} 2.020} \\
W$_{Ti}$                 & {\color[HTML]{3531FF} 3.878} & {\color[HTML]{FE0000} 9.759} & {\color[HTML]{3531FF} 2.027} & {\color[HTML]{3531FF} 2.027} \\
Cr$_{Ti}$                & {\color[HTML]{FE0000} 3.876} & {\color[HTML]{FE0000} 9.743} & {\color[HTML]{FE0000} 1.987} & {\color[HTML]{FE0000} 1.929} \\
La$_{Ti}$                & {\color[HTML]{3531FF} 3.874} & {\color[HTML]{3531FF} 9.990} & {\color[HTML]{3531FF} 2.375} & {\color[HTML]{3531FF} 2.356} \\
Cu$_{Ti}$                & {\color[HTML]{3531FF} 3.873} & {\color[HTML]{FE0000} 9.727} & {\color[HTML]{3531FF} 2.031} & {\color[HTML]{3531FF} 1.986} \\
Co$_{Ti}$                & {\color[HTML]{FE0000} 3.868} & {\color[HTML]{FE0000} 9.736} & {\color[HTML]{FE0000} 1.943} & {\color[HTML]{FE0000} 1.944} \\
Ce$_{Ti}$                & {\color[HTML]{3531FF} 3.885} & {\color[HTML]{3531FF} 9.838} & {\color[HTML]{3531FF} 2.292} & {\color[HTML]{3531FF} 2.164} \\ \bottomrule
\end{tabular}
\end{center}
The values have been color coded. Blue corresponds to increment and red corresponds to decrement with respect to parameters of pristine anatase.
\end{table}

\section{Band structures}
\begin{figure}[htbp]
     \centering
         \includegraphics[width=0.98\linewidth]{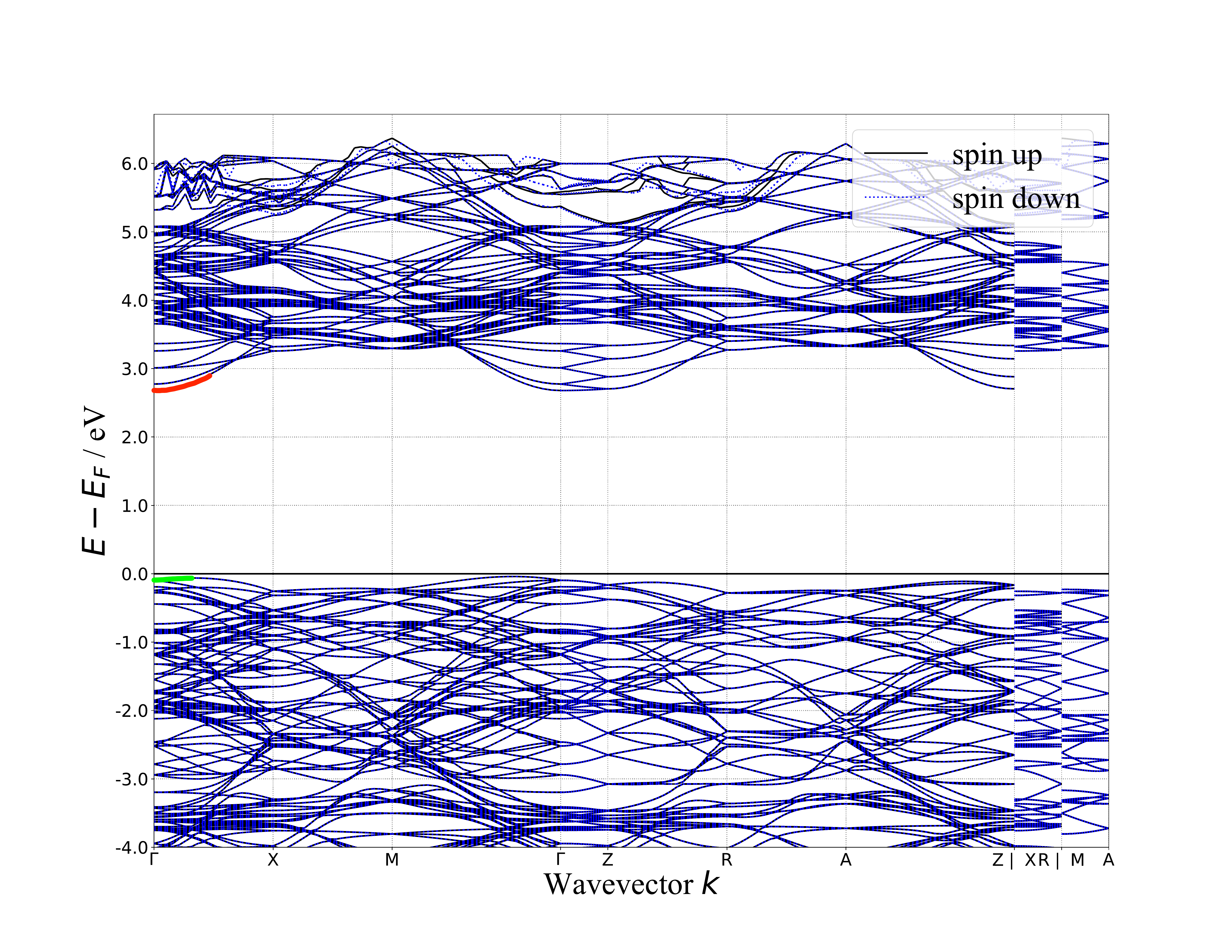}
     \caption{Band structure of pristine anatase. Fermi energy has been shifted to 0 eV.
     The band segments used for band dispersion fitting along the $\Gamma-X$ direction have been marked for the conduction band (red) and for the valence band (green).
     The extrema of the conduction band segment lies at the $\Gamma$ point while in case of the valence band segment, the extrema point lies at the right end of the segment.
     }
     \label{SUP:1}
\end{figure}
\begin{figure}[htbp]
     \centering
         \includegraphics[width=0.98\linewidth]{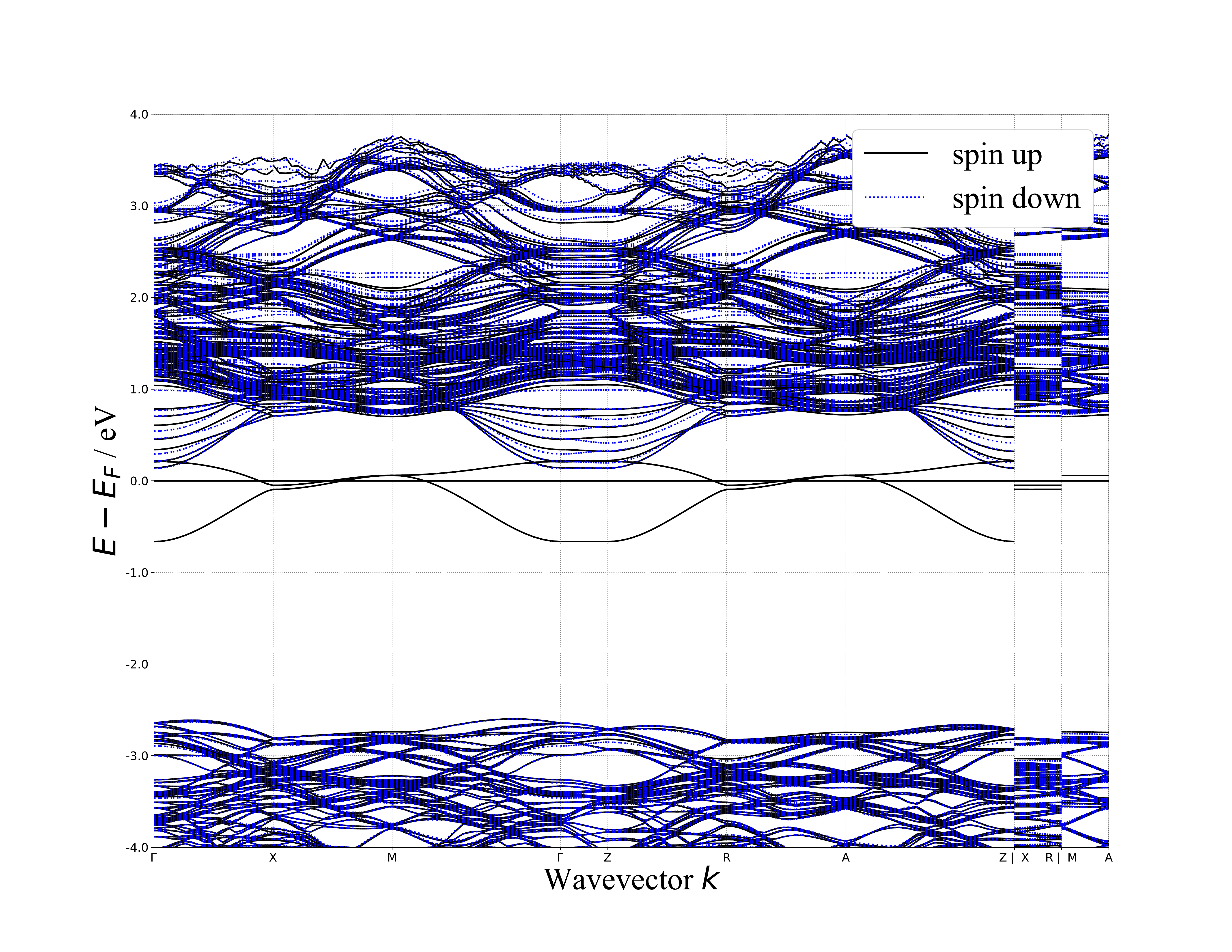}
     \caption{Band structure of Nb-doped anatase. Fermi energy has been shifted to 0 eV.}
     \label{SUP:2}
\end{figure}
\begin{figure}[htbp]
     \centering
         \includegraphics[width=0.98\linewidth]{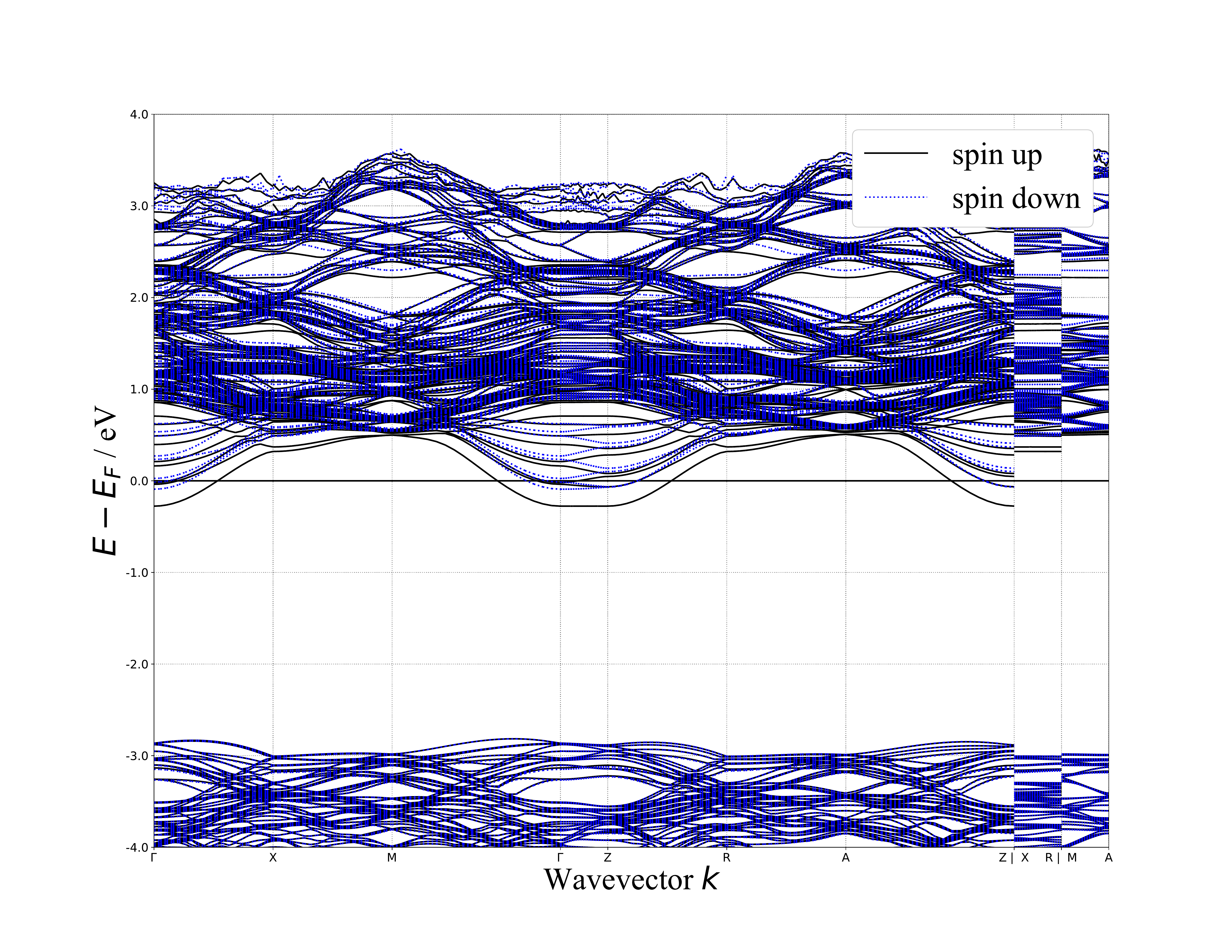}
     \caption{Band structure of Ta-doped anatase. Fermi energy has been shifted to 0 eV.}
     \label{SUP:3}
\end{figure}
\begin{figure}[htbp]
     \centering
         \includegraphics[width=0.98\linewidth]{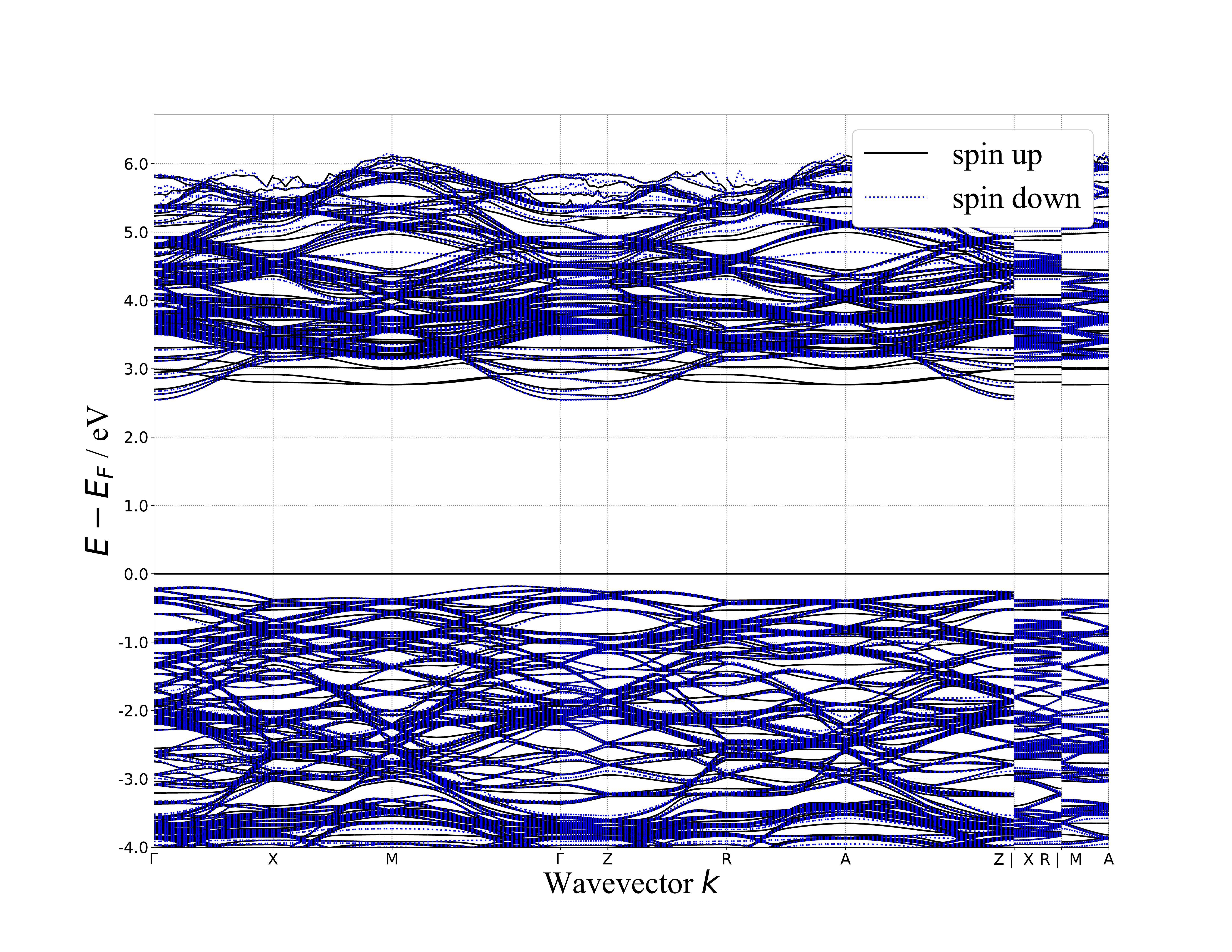}
     \caption{Band structure of V-doped anatase. Fermi energy has been shifted to 0 eV.}
     \label{SUP:4}
\end{figure}
\begin{figure}[htbp]
     \centering
         \includegraphics[width=0.98\linewidth]{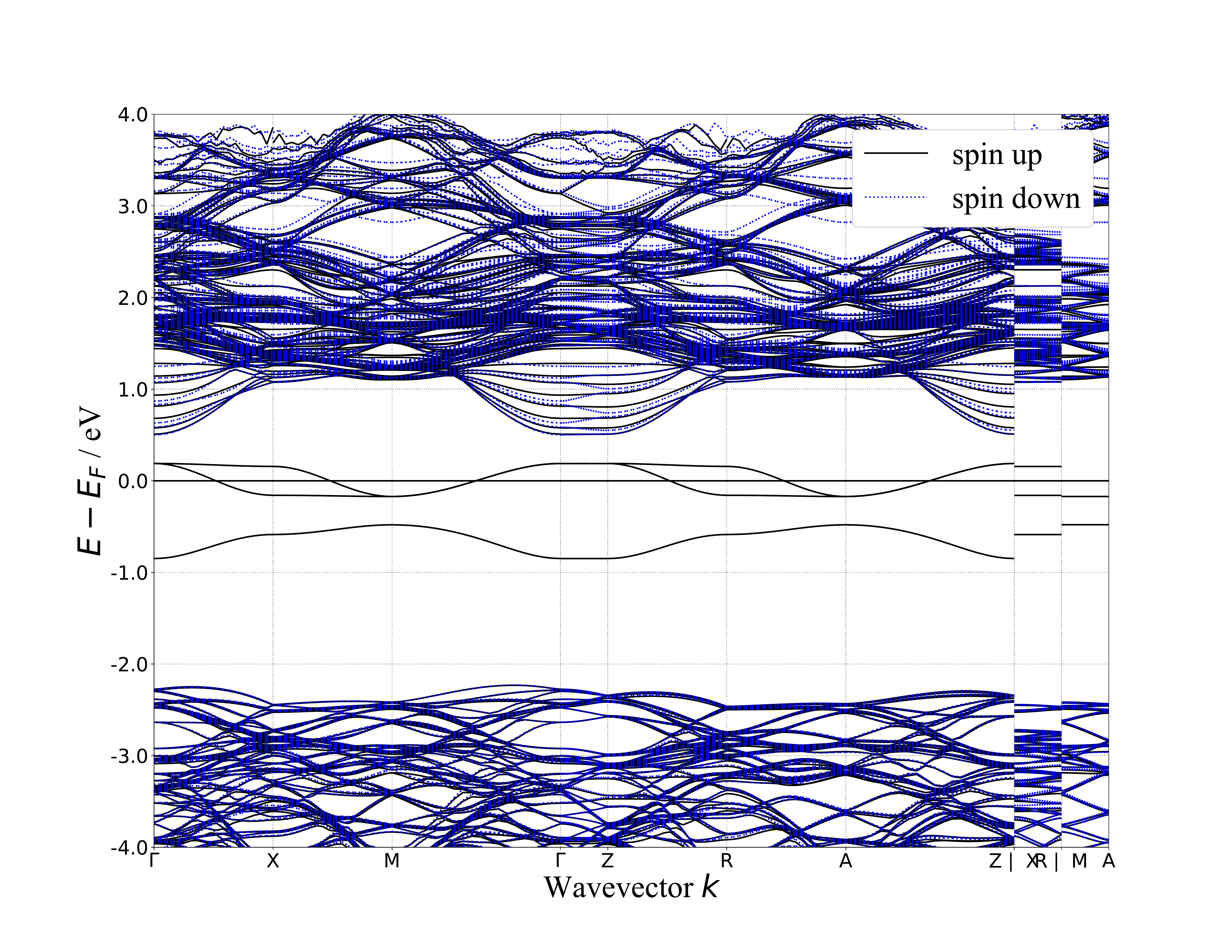}
     \caption{Band structure of Mo-doped anatase. Fermi energy has been shifted to 0 eV.}
     \label{SUP:5}
\end{figure}
\begin{figure}[htbp]
     \centering
         \includegraphics[width=0.98\linewidth]{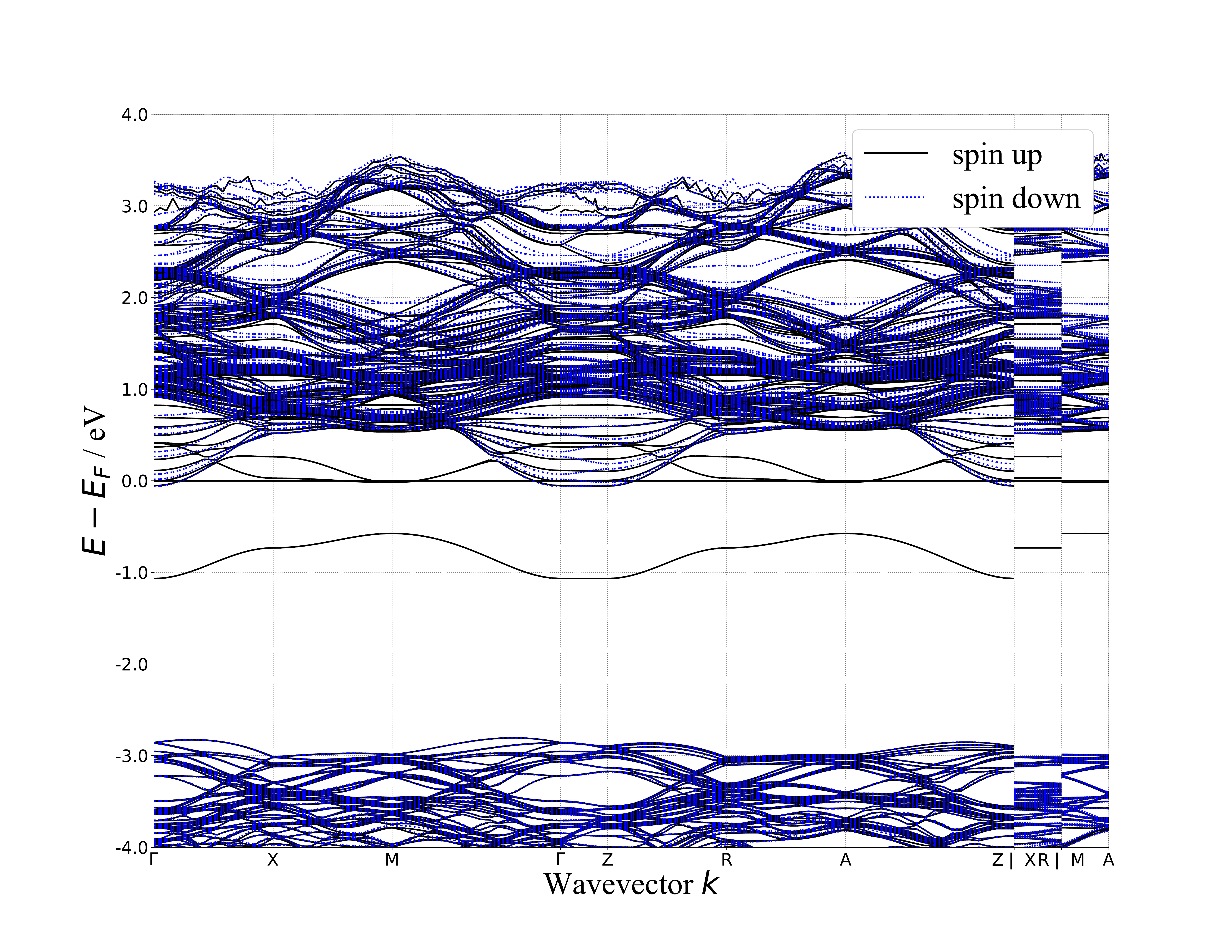}
     \caption{Band structure of W-doped anatase. Fermi energy has been shifted to 0 eV.}
     \label{SUP:6}
\end{figure}
\begin{figure}[htbp]
     \centering
         \includegraphics[width=0.98\linewidth]{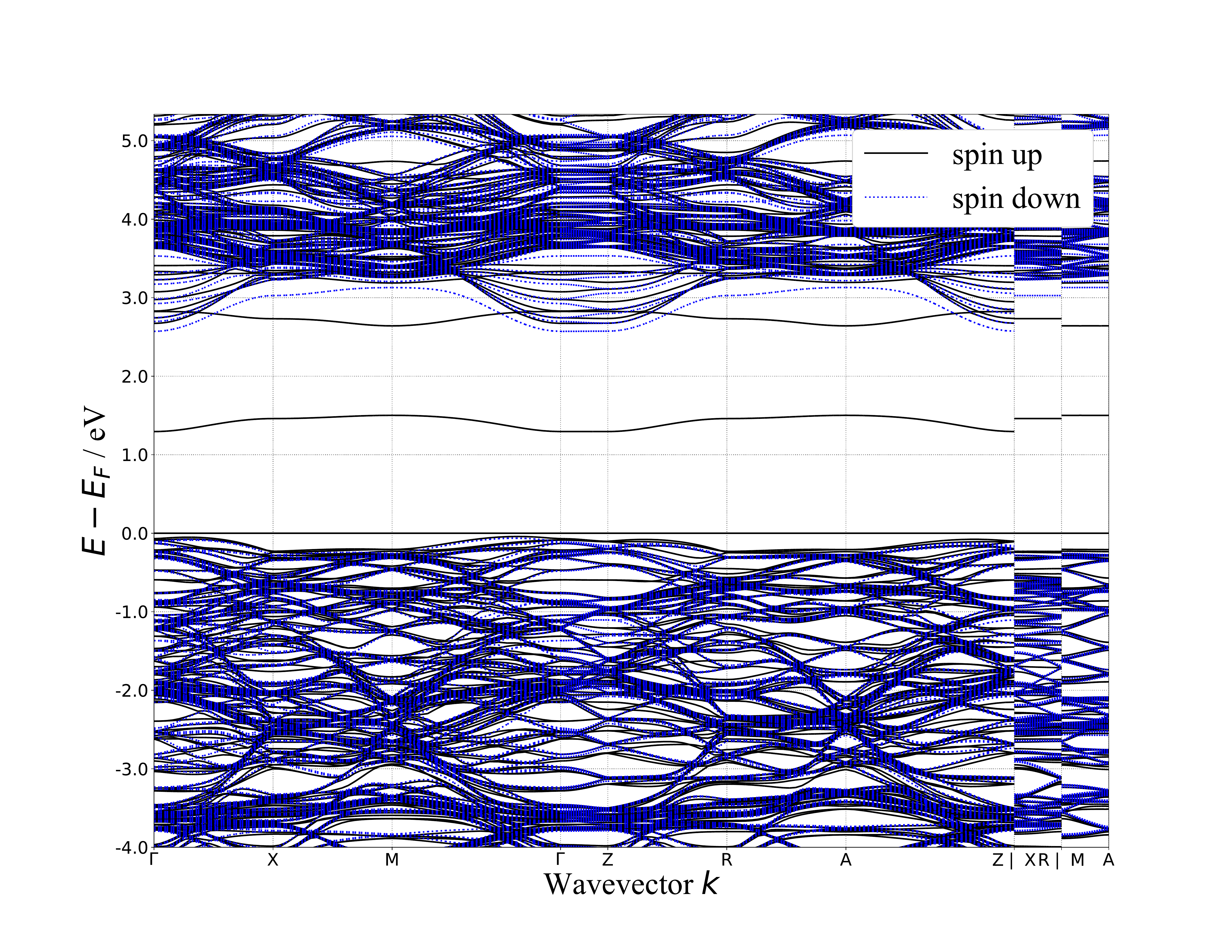}
     \caption{Band structure of Cr-doped anatase. Fermi energy has been shifted to 0 eV.}
     \label{SUP:7}
\end{figure}
\begin{figure}[htbp]
     \centering
         \includegraphics[width=0.98\linewidth]{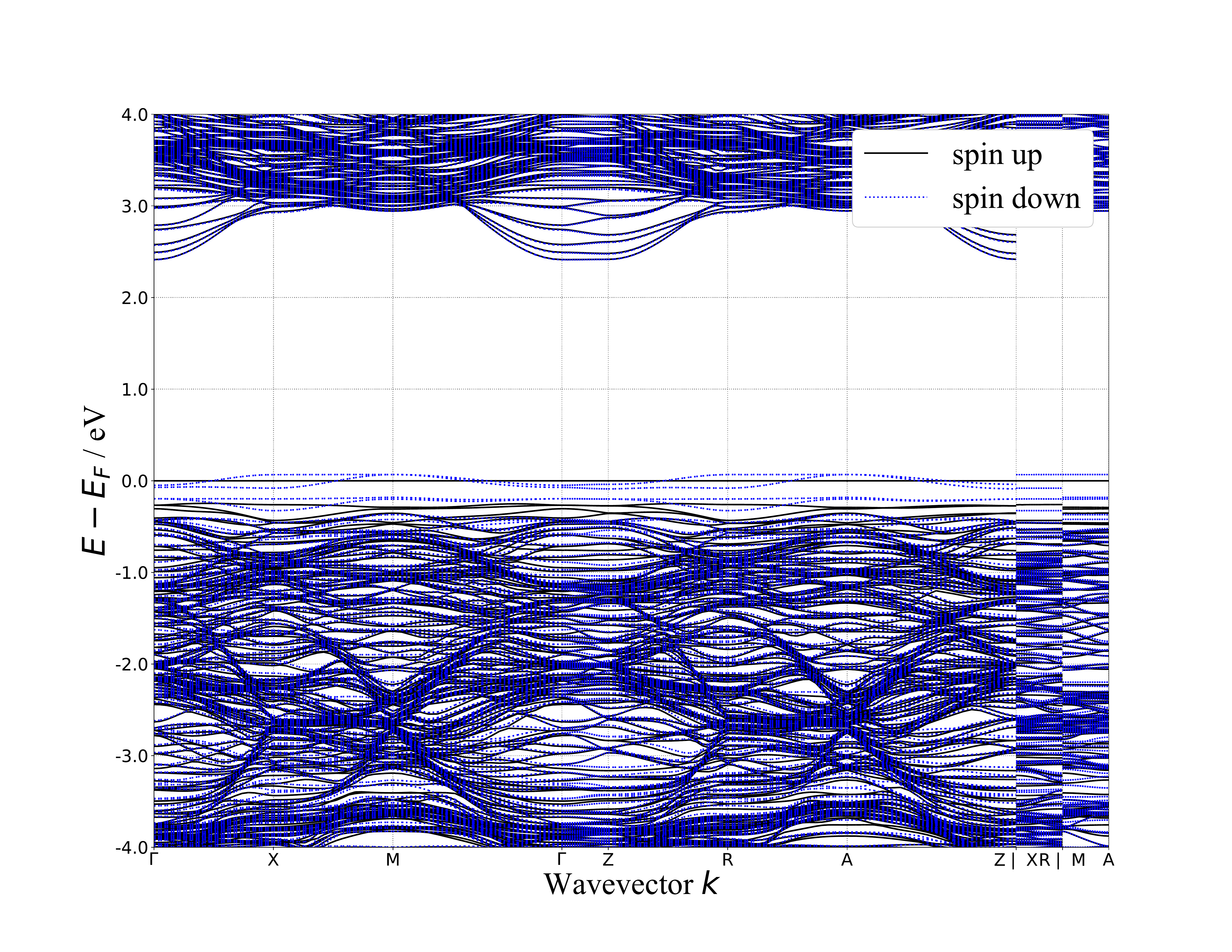}
     \caption{Band structure of La-doped anatase. Fermi energy has been shifted to 0 eV.}
     \label{SUP:8}
\end{figure}
\begin{figure}[htbp]
     \centering
         \includegraphics[width=0.98\linewidth]{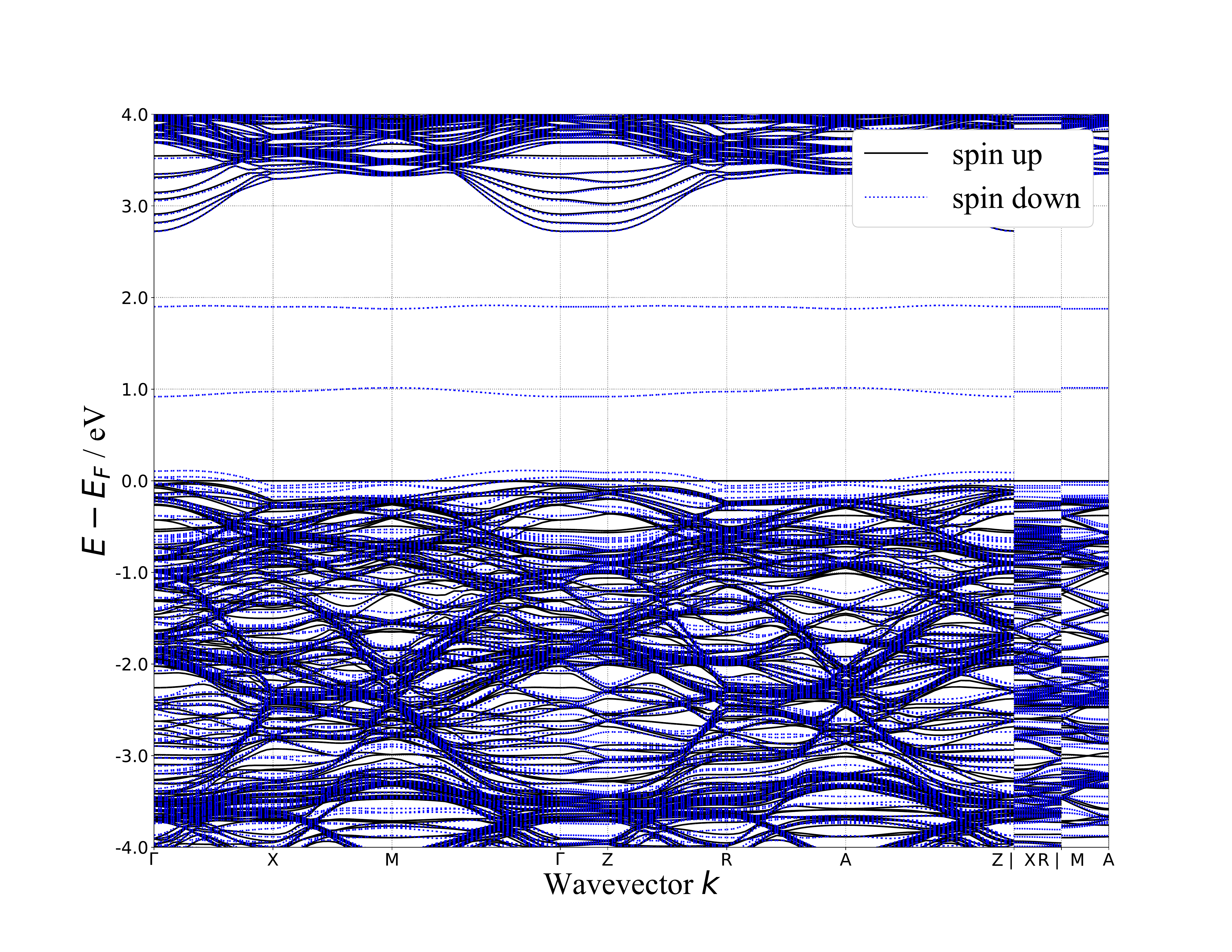}
     \caption{Band structure of Cu-doped anatase. Fermi energy has been shifted to 0 eV.}
     \label{SUP:9}
\end{figure}
\begin{figure}[htbp]
     \centering
         \includegraphics[width=0.98\linewidth]{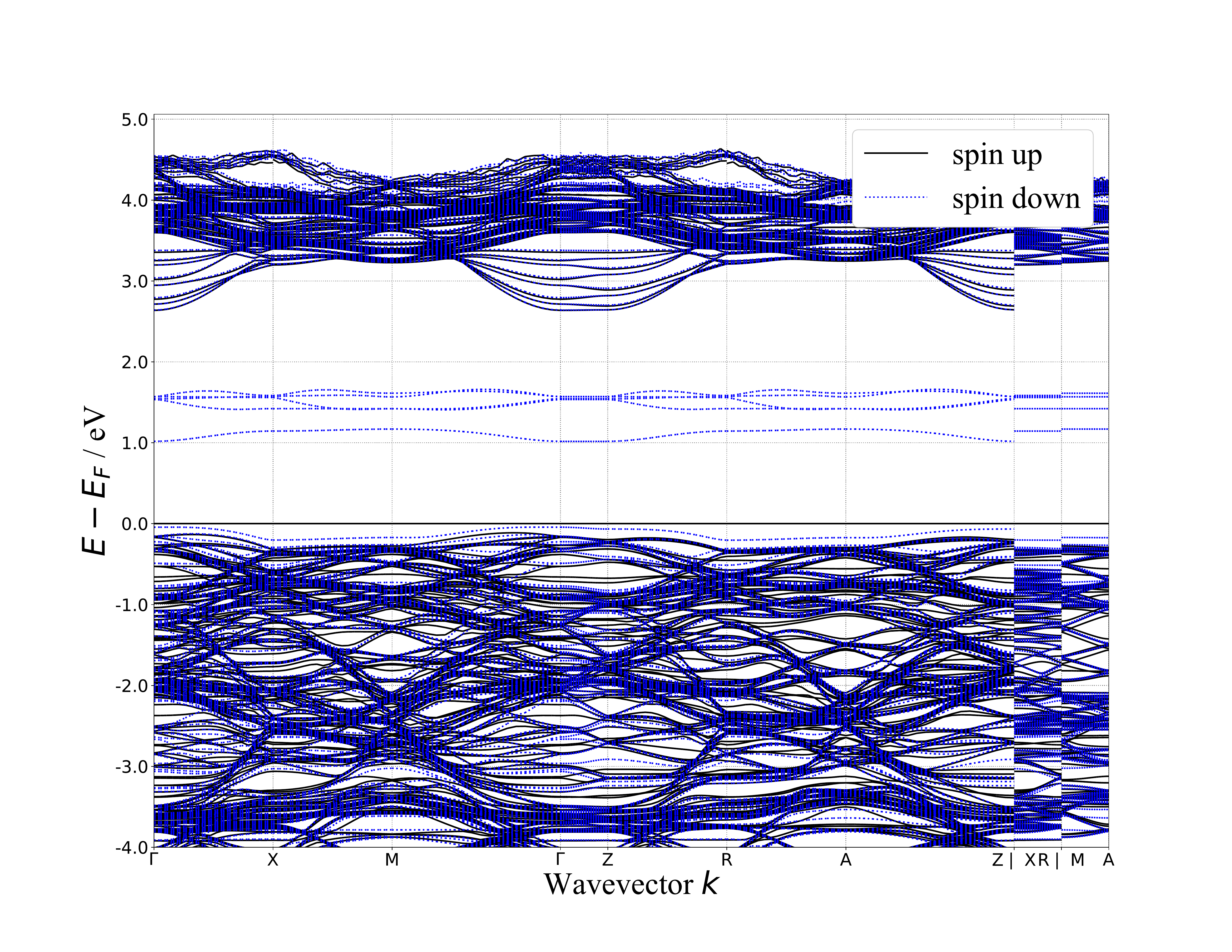}
     \caption{Band structure of Co-doped anatase. Fermi energy has been shifted to 0 eV.}
     \label{SUP:10}
\end{figure}
\begin{figure}[htbp]
     \centering
         \includegraphics[width=0.98\linewidth]{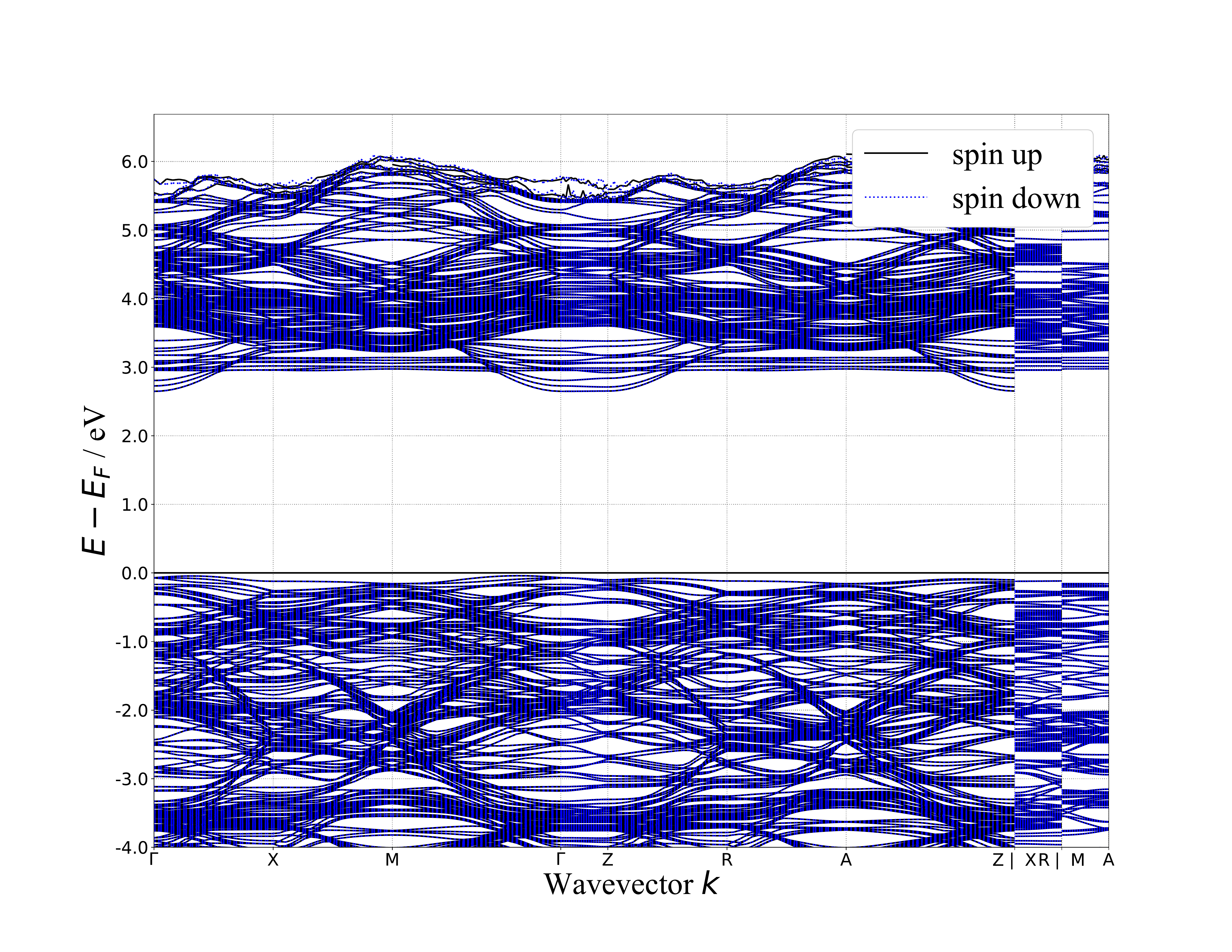}
     \caption{Band structure of Ce-doped anatase. Fermi energy has been shifted to 0 eV.}
     \label{SUP:11}
\end{figure}